\documentclass[a4paper,12pt]{article}
\usepackage{ulem}
\usepackage{subcaption}

\usepackage[margin=1in]{geometry}    
\usepackage{graphicx}               
\usepackage{sectsty}                
\usepackage{fancyhdr}               
\usepackage{hyperref}               
\usepackage{xcolor}                 
\usepackage{listings}               
\usepackage{enumitem}               
\usepackage{caption}                
\usepackage{tikz}                   
\usepackage{float}
\usepackage{array}                  
\usepackage{xcolor}                 
\usepackage{caption}                
\usepackage{booktabs}               
\usepackage{tabularx}               
\usepackage{makecell} 
\usepackage{listings}
\usepackage{graphicx} 
\usepackage{rotating} 
\usepackage[export]{adjustbox} 
\usepackage{atbegshi}
\usepackage{listings}
\hypersetup{
    colorlinks=true,
    linkcolor=blue,
    filecolor=magenta,
    urlcolor=cyan,
}

\sectionfont{\color{blue}\bfseries\Large}
\subsectionfont{\color{blue}\bfseries\normalsize}
\subsubsectionfont{\color{blue}\bfseries\small}
\newcolumntype{C}[1]{$>$ {\centering\textbackslash}m{#1}}  

\AtBeginShipout{%
  \AtBeginShipoutAddToBoxForeground{%
    \begin{tikzpicture}[remember picture,overlay]
      \node[opacity=0.15, rotate=45] at (current page.center) {
           \scalebox{8}{\textbf{\textcolor{black}{}}} 
      };
    \end{tikzpicture}%
  }%
}

\lstdefinelanguage{yaml}{
  keywords={true, false, null, yes, no, description, title, logsource, detection, condition, fields, tags, falsepositives, mitre, author, date, id, status},
  keywordstyle=\color{blue}\bfseries,
  basicstyle=\ttfamily\small,        
  comment=[l]{\#},                   
  commentstyle=\color{gray}\itshape, 
  morestring=[b]',                   
  morestring=[b]",                   
  stringstyle=\color{red},           
}

\begin{document}

\begin{titlepage}
        \centering
    \vspace*{1cm}
    {\Large\textcolor{black}{\textbf{Under the Hood of BlotchyQuasar: DLL-Based RAT Campaigns Against Latin America}}} \\[1cm]
    Alessio Di Santo (alessio.disanto@graduate.univaq.it)\\
    Università degli Studi dell’Aquila, L’Aquila, Abruzzo, Italy  \\
    \textbf{Date:} August 24, 2024 \\
    \vfill
    {\Large\textcolor{gray!60}{\it "Non videmus ea quae mox futura sunt"}} \\[0.8cm]
    {\small\textcolor{gray!60}{(We do not see the things that will soon be) — Marcus Tullius Cicero}} \\
    \vfill
\end{titlepage}

\tableofcontents
\newpage

\section{Executive Summary}
A sophisticated malspam campaign was uncovered targeting Latin American countries, with a particular focus on Brazil, on August 19, 2024. This operation utilizes a highly deceptive phishing email to trick users into executing a malicious \textit{MSI} file, initiating a multi-stage infection. The core of the attack leverages \textit{DLL} side-loading, where a legitimate executable from Valve Corporation is used to load a trojanized \textit{DLL}, thereby bypassing standard security defenses.

Once active, the malware, a variant of \textit{QuasarRAT} known as \textit{BlotchyQuasar}, is capable of a wide range of malicious activities. It is designed to steal sensitive browser-stored credentials and banking information, the latter through fake login windows mimicking well-known \textit{Brazilian} banks. The threat establishes persistence by modifying the Windows registry , captures user keystrokes through keylogging , and exfiltrates stolen data to a \textit{Command-and-Control} server using encrypted payloads. Despite its advanced capabilities, the malware’s code exhibits signs of rushed development, with inefficiencies and poor error handling that suggest the threat actors prioritized rapid deployment over meticulous design. Nonetheless, the campaign’s extensive reach and sophisticated mechanisms pose a serious and immediate threat to the targeted regions, underscoring the need for robust cybersecurity defenses.
\newpage

\section{Introduction}
\subsection{Objective}
The objective of this \textit{Malware Analysis Report} is to provide an in-depth understanding of the behavior, architecture, and intent of a malicious software instance. At its core, this report serves as a crucial tool for identifying the characteristics and operations of the \textit{threat}, offering detailed insights that can be used to map the broader attack landscape. By dissecting the capabilities and infrastructure of the malware, analysts are able to build a clear picture of its functionality, origin, and potential impact.

Mapping a \textit{threat} accurately is of paramount importance for defenders. A well-crafted malware analysis report helps connect individual malicious artifacts with broader attack campaigns and identifies common \textit{Techniques, Tactics, and Procedures} (\textit{TTPs}) employed by adversaries. This intelligence feeds into a larger knowledge base that allows cybersecurity teams to understand how threats evolve, recognize new campaigns with similar signatures, and anticipate potential next steps of attackers. The report is not merely an exercise in detailing technical specifics but also a way of enriching the collective understanding of a \textit{Threat Actor}'s capabilities, motivations, and behaviors.

Actionable \textit{Threat Intelligence} derived from malware analysis is particularly valuable because it enables proactive defenses. With a structured understanding of the malware’s \textit{Indicators of Compromise} (\textit{IOCs}), behavioral patterns, and infrastructure, \textit{Threat Hunting} and \textit{Monitoring} teams are equipped with the context needed to seek out malicious activity before it fully manifests. \textit{Threat Hunters} can leverage this intelligence to identify adversarial presence across their environments more effectively, while \textit{Monitoring} teams can enhance detection logic and fine-tune alerts to identify these threats more accurately in real time. This coordinated approach bolsters an organization’s defense posture, making it possible to detect and respond to even well-structured, sophisticated threats that are designed to evade traditional security mechanisms.

Ultimately, a comprehensive malware analysis report provides not only a retrospective view of what a threat has done but also equips defenders with the tools and knowledge to better \textit{predict}, \textit{detect}, and \textit{prevent} future attacks. This knowledge empowers security teams to make informed decisions, prioritize vulnerabilities, and improve their capabilities against \textit{Advanced Persistent Threats} (\textit{APTs}).

\subsection{Infection Chain}

\begin{figure}[H]
    \centering
    \frame{\includegraphics[width=1\linewidth,frame]{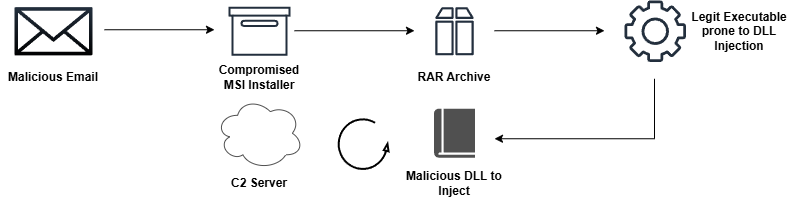}}
    \caption{Infection Chain Diagram}
    \label{fig:00}
\end{figure}

\newpage

\section{Methodology}
Analyzing the malware involved a comprehensive approach utilizing both static and dynamic analysis techniques to thoroughly understand its structure, behavior, and potential impact. By combining these two approaches, it is possible to gain a comprehensive understanding of the malware's capabilities and objectives. Static analysis provided insights into its structure and obfuscation methods, while dynamic analysis revealed its real-time behavior and interactions with the system. This dual approach was essential in developing effective detection and mitigation strategies against this sophisticated threat.
\subsection{Static Analysis}
Static analysis is a fundamental technique in malware analysis that involves examining the code of malicious software without executing it. This approach focuses on understanding the structure, logic, and intent of the malware through methods such as \textit{disassembling}, \textit{decompiling}, and reviewing its binary or script content. By analyzing the static properties of malware, such as strings, embedded resources, file headers, and imported functions, researchers can gather valuable insights into its capabilities, communication patterns, and potential targets.

The main goal of static analysis is to dissect the malware's inner workings, identify hardcoded \textit{Indicators of Compromise} (\textit{IoCs}) like IP addresses, URLs, or file paths, and infer its behavior without the risk of executing harmful code. This method is particularly useful for uncovering obfuscation techniques, encrypted payloads, and multi-stage architectures, which are often employed by modern malware to hinder direct analysis.

However, static analysis comes with its challenges. Advanced malware frequently uses obfuscation, packing, or encryption to conceal its code and deter examination. Analysts must rely on specialized tools and techniques, such as deobfuscation scripts, unpackers, and cryptographic analysis, to overcome these barriers. Moreover, analyzing assembly-level or machine code demands a high level of expertise, as the complexity of the malware's logic can obscure its true intent.

Despite its limitations, static analysis is invaluable as it allows analysts to preemptively assess a malware sample’s potential threats, providing critical intelligence without the inherent risks of execution. Combined with dynamic analysis, it forms a comprehensive approach to malware investigation, equipping defenders with the necessary understanding to develop effective detection and mitigation strategies.

\subsection{Dynamic Analysis}
Dynamic analysis is a cornerstone of malware analysis, enabling researchers to observe the behavior of malicious software in real-time by executing it within a controlled, isolated environment. This approach is particularly valuable for analyzing modern malware that employs sophisticated \textit{obfuscation techniques}, rendering static analysis alone insufficient. By simulating realistic conditions, analysts can examine how malware interacts with the file system, registry, processes, network, and system \textit{API}s, providing direct insights into its functionality and intent.

The objective of dynamic analysis is to uncover the behavioral profile of the malware, revealing actions such as \textit{data exfiltration}, \textit{Command-and-Control} communication, \textit{credential theft}, and \textit{persistence mechanisms}. It also aids in identifying \textit{Indicators of Compromise} (\textit{IoCs}), such as IP addresses, domains, and modified system configurations, which are crucial for detection and response efforts. This method is not without challenges, as modern malware often incorporates \textit{anti-analysis techniques} designed to detect and evade \textit{Sandboxed Environments}, \textit{Virtual Machines}, or \textit{Debugging Tools}. These measures include delaying execution, checking for artifacts indicative of analysis environments, and employing runtime obfuscation to conceal its activities.

Despite these difficulties, dynamic analysis remains a critical tool in the fight against advanced threats. Its ability to reveal runtime behavior complements static analysis, providing a comprehensive understanding of the malware’s objectives and capabilities. While the process can be resource-intensive and time-consuming, its contributions to cybersecurity are indispensable, offering valuable intelligence to counteract and mitigate malicious campaigns effectively.
\newpage
\section{Quasar RAT a.k.a. BlotchyQuasar}
\textbf{\textit{QuasarRAT}}, a potent and versatile \textit{Remote Access Trojan}, presents a significant and evolving threat in the contemporary cybersecurity landscape. Its open-source nature, originally intended for legitimate administrative purposes, has been twisted by a wide array of malicious actors, from sophisticated state-sponsored groups to individual cybercriminals. This analysis will explore the technical capabilities of \textit{QuasarRAT}, delve into the profile of its creator and the dichotomy of its intended use versus its nefarious applications, identify the prominent threat actors who employ it, and examine some of the most notorious campaigns where it has played a central role\footnote{https://malpedia.caad.fkie.fraunhofer.de/details/win.quasar\_rat}.

Developed in \textit{C\#} and first appearing around 2014, \textit{QuasarRAT} offers a comprehensive suite of functionalities that make it an attractive tool for remote system management. These capabilities include, but are not limited to, keylogging to capture sensitive user input, comprehensive file system access for data exfiltration and manipulation, the ability to establish a reverse proxy for covert communication, and the power to execute arbitrary code on a compromised machine. A key technical characteristic that enhances its stealth is the adept use of \textit{Dynamic-Link Library} (\textit{DLL}) side-loading. This technique allows the malware to load malicious \textit{DLL}s by exploiting how legitimate applications search for and load these libraries, effectively camouflaging its presence and evading detection by traditional security solutions. This method, combined with encrypted \textit{C2} communication, makes \textit{QuasarRAT} a formidable and difficult-to-detect threat.

The creator of \textit{QuasarRAT}, who operated under the alias \textit{quasar}, originally developed the tool as a legitimate, open-source remote administration tool. The intention was to provide system administrators with a free and powerful utility to manage their networks. This benevolent origin, however, has been completely overshadowed by its widespread adoption by the malicious community. The very features that make it a useful administrative tool—remote access, file transfer, and process management—are precisely what make it an effective weapon in the hands of attackers. This dual-use nature highlights a persistent challenge in the software world, where powerful tools can be easily repurposed for illicit activities. The open-source availability of its code on platforms like \textit{GitHub} further exacerbates this issue, as it allows threat actors to freely modify and customize the RAT to suit their specific operational needs and to develop new variants that can bypass updated security measures\footnote{https://insights.bridewell.com/hubfs/Cyber\%20Threat\%20Intelligence\%20Report\%202025.pdf}.

The spectrum of threat actors employing \textit{QuasarRAT} is broad and varied. At one end are highly sophisticated \textit{Advanced Persistent Threat} (\textit{APT}) groups, often with ties to nation-states. A prominent example is \textit{APT10}, also known as \textit{Cicada} or \textit{Stone Panda}, a Chinese-linked group that has been observed using \textit{QuasarRAT} in numerous campaigns targeting a wide range of sectors, including government, defense, and technology, across the globe. For these actors, \textit{QuasarRAT} serves as a reliable and customizable tool for espionage, data theft, and establishing a long-term foothold within target networks. Another notable state-sponsored actor is the \textit{Kimsuky} group, believed to be of North Korean origin, which has utilized \textit{QuasarRAT} in its intelligence-gathering operations.

Beyond nation-state actors, this threat is also a favored tool among cybercriminal organizations and individual hackers. Its ease of use and readily available documentation lower the barrier to entry for less sophisticated attackers. These groups often deploy \textit{QuasarRAT} in phishing campaigns to gain initial access to systems, with the ultimate goal of financial gain through ransomware deployment, credential theft, or the sale of stolen data on underground forums. The versatility of the \textit{RAT} allows for a range of criminal activities, making it a persistent threat to businesses and individuals alike.

Several high-profile cases underscore the significant impact of \textit{QuasarRAT}. In numerous instances, it has been a key component in campaigns targeting critical infrastructure and government entities. For example, threat actors have leveraged this tradecraft in attacks against \textit{Ukrainian} organizations, often delivered through spear-phishing emails containing malicious attachments\footnote{https://socprime.com/it/blog/uac-0050-attack-detection-hackers-are-armed-with-remcos-rat-quasar-rat-and-remote-utilities-to-target-ukraine-once-again/}. These campaigns aim to disrupt operations and exfiltrate sensitive information. The use of \textit{QuasarRAT} in these contexts demonstrates its effectiveness as a tool for cyber warfare and political espionage.

Furthermore, the evolution of \textit{QuasarRAT}'s deployment tactics is evident in various campaigns. Threat actors have been observed combining this \textit{RAT} with other malware and tools to create more complex and effective attack chains. For instance, it has been used in conjunction with information stealers to maximize data harvesting from compromised systems. The continuous development of new variants and delivery methods by malicious actors ensures that \textit{QuasarRAT} remains a relevant and dangerous threat that requires constant vigilance and advanced threat detection capabilities to mitigate.

In conclusion, \textit{QuasarRAT} exemplifies the dual-edged sword of open-source software. While born from a legitimate desire to create a useful administrative tool, its powerful features and accessibility have made it a go-to weapon for a wide range of threat actors. Its technical sophistication, particularly its stealth capabilities, combined with its ease of customization, ensures its continued prevalence in the threat landscape. The notorious campaigns in which it has been a key tool serve as a stark reminder of the significant damage that can be inflicted by this versatile remote access trojan, making it a critical threat for organizations to understand and defend against.

\newpage
\section{Analysis Results}
\subsection{C2 Infrastructure}
\begin{figure}[H]
    \centering
    \frame{\includegraphics[width=0.8\linewidth]{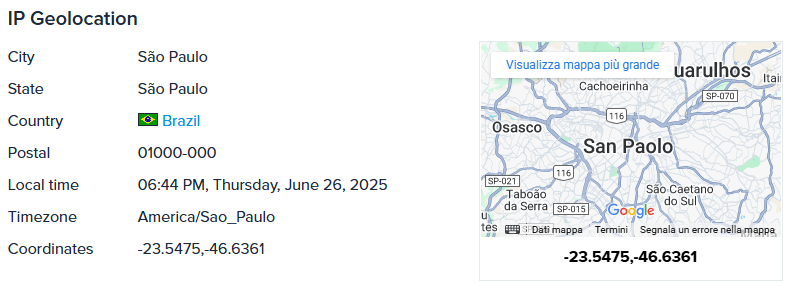}}
    \caption{Geographical location of the subjected server.}
    \label{fig:geo}
\end{figure}

Upon close examination, the IP address \textit{15.228.186[.]93} reveals itself to be a significant node in a malicious cyber infrastructure, actively engaged in the distribution of malware. A deep dive into its activity and context provides a clear picture of its role in ongoing cyber threats. The address is geolocated to \textit{São Paulo}, Brazil, and is part of the \textit{AS16509 Autonomous System}, which belongs to \textit{Amazon Data Services Brazil} (Fig. \ref{fig:geo}). The decision by threat actors to operate from within a major cloud provider's network, such as \textit{AWS}, is a deliberate and common tactic. It allows them to leverage the provider's reputable and vast IP space, complicating efforts by network defenders to block their activities without risking the interruption of legitimate cloud-hosted services.

The primary function of this IP appears to be the hosting and dissemination of malicious payloads. Open-source intelligence and data from malware repositories like \textit{MalwareBazaar} and \textit{ANY.RUN} have directly implicated \textit{15.228.186[.]93} in these activities. For instance, it was observed serving as a direct download source for a malicious loader, specifically retrieving a file named \textit{Rar.exe} from a URL constructed with the IP address.

Further investigation, corroborated by threat intelligence feeds such as \textit{Maltrail}, firmly associates the IP with the \textit{Command-and-Control} (\textit{C2}) infrastructure for \textit{QuasarRAT}. The latter is a potent and well-known open-source \textit{Remote Access Trojan} (\textit{RAT}) built on the \textit{.NET framework}. Its widespread use stems from its extensive feature set, which includes capabilities for keylogging, credential theft from browsers and applications, screen capturing, establishing a reverse proxy for covert network access, and comprehensive file system manipulation. The open-source nature of \textit{QuasarRAT} enables various malicious actors to easily adapt and deploy it for their specific campaigns.

The malicious infrastructure extends beyond the single IP address, incorporating a network of domains that resolve to \textit{15.228.186[.]93}. These domains, including \textit{agicltursement[.]ink}, \textit{cfestlolequiep[.]store}, \textit{gastronomleo[.]lat}, \textit{mercantokiko[.]xyz}, \textit{noticiasnovidads[.]xyz}, \textit{varjolatijolos[.]space}, and the dynamic \textit{DNS hostname} \textit{coletasegura[.]ddns[.\newline]net}, form the functional backbone of the \textit{C2} communications. The use of varied, often newly registered, top-level domains alongside \textit{dynamic DNS services} is a classic strategy to build a resilient and evasive \textit{C2} network, allowing for rapid changes to evade blacklisting.

The operational \textit{Tactics, Techniques and Procedures} (\textit{TTPs}) employed by the operators of this infrastructure follow a familiar pattern. An attack would typically commence with an initial access vector like a phishing email, which persuades a user to execute the initial loader. Once running, this downloader establishes contact with the \textit{C2} server at \textit{15.228.186[.]93} to pull down the next stage of the attack, in this case, the \textit{QuasarRAT} payload. The malware then entrenches itself on the victim's system, leveraging defense evasion techniques while it establishes a persistent, encrypted channel back to the \textit{C2} server. Through this channel, the attacker gains full remote control over the compromised machine, enabling data exfiltration, surveillance, and the potential to use the machine as a pivot point for further attacks within the network. Given these activities, it is imperative for network defenders to block this IP address and its associated domains at their network perimeters, hunt for the identified malware hash within their environments, and maintain vigilant security practices to mitigate the risk posed by such threats.

\subsection{Dissecting the attack}
\begin{figure}
    \centering
    \frame{\includegraphics[width=0.7\linewidth]{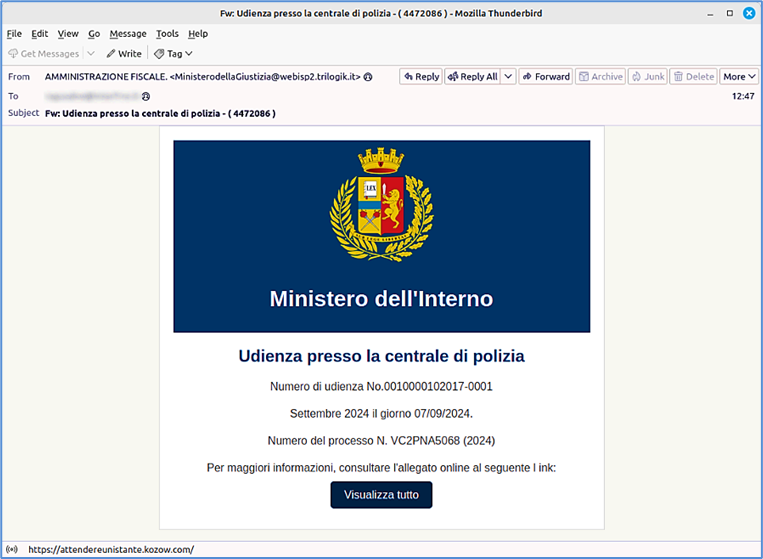}}
    \caption{Malspam E-mail sent to Italian users.}
    \label{fig:malemeail}
\end{figure}

On August 19, 2024, a malspam campaign titled \textit{Fw: Informazioni sulle entrate governative. - (2607579)}\footnote{https://cert-agid.gov.it/news/cresce-lattivita-di-quasar-rat-blotchyquasar-contro-gli-utenti-di-istituti-bancari-italiani/} was distributed to an undisclosed number of Italian recipients (Fig. \ref{fig:malemeail}). Concurrently, a similar campaign was detected, primarily targeting users in \textit{Latin American} countries (including \textit{Argentina}, \textit{Brazil}, \textit{Mexico}, \textit{Chile}, \textit{Colombia}, \textit{Costa Rica}, \textit{Panama}, \textit{Ecuador}, \textit{Peru}, \textit{Uruguay}, and \textit{Venezuela}), with a particular emphasis on \textit{Brazilian} users. The infection chain is initiated through an email, which deceives the recipient into downloading a malicious \textit{MSI} file under the pretense of viewing non-existent documents. In the Italian version, for instance, the email falsely suggests that the victim needs to review evidence as part of an upcoming trial in which they are involved.

\begin{figure}[H]
    \centering
    \frame{\includegraphics[width=0.8\linewidth]{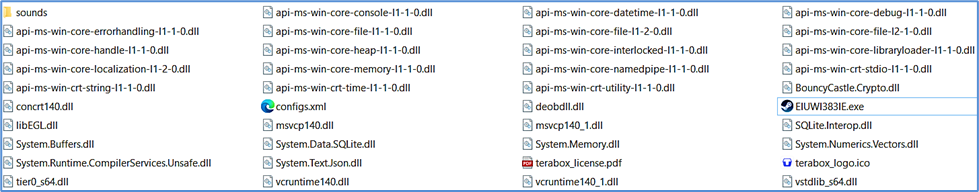}}
    \caption{Content embedded inside \textit{xxwewe33.rar} archive.}
    \label{fig:rararch}
\end{figure}

Both \textit{Italian} and \textit{LATAM} users were than tricked to download a malicious \textit{MSI} file. However, for Latin American victims, the installer \textit{67dee1.msi} is subsequently downloaded and, when executed, attempts to connect to the following URLs to retrieve both a malicious archive (Fig. \ref{fig:rararch}) and a legitimate instance of \textit{WINRAR}:
\begin{itemize}
    \item http://15.228.186[.]93/33354365346/Rar.exe
    \item http://15.228.186[.]93/33354365346/xxwewe33.rar
\end{itemize}

Once both these files were downloaded, \textit{eiuwi383ie.exe} was executed.

To gain a comprehensive understanding of this threat's capabilities, several analyses were conducted on the contents of the \textit{xxwewe33.rar} archive. Initially, the analysis revealed the presence of both a \textit{PDF} and an \textit{ICO} file associated with \textit{Terabox}, a document-sharing platform. This suggests that the attacker attempted to obscure the true nature of the executable by including misleading elements, such as fake evidence pointing to a \textit{Terabox} instance, which could be perceived as a legitimate tool for viewing trial-related documentation.

\begin{figure}[H]
    \centering
    \frame{\includegraphics[width=0.8\linewidth]{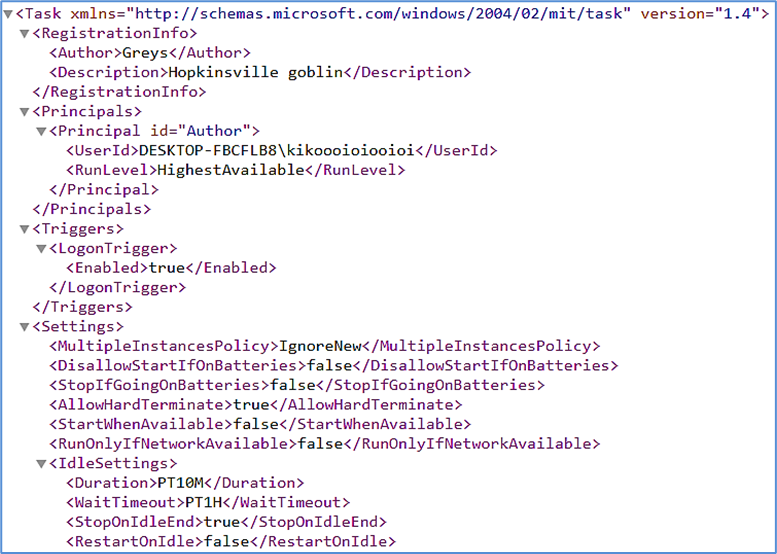}}
    \caption{\textit{config.xml} information snippet.}
    \label{fig:config}
\end{figure}

Further analyses on the available \textit{config.xml} file (Fig. \ref{fig:config}) allows to gather the following useful Intelligence:
\begin{itemize}
    \item $<$ Author$>$ \textbf{Greys} $<$/Author$>$
    \item $<$ Description$>$ \textbf{Hopkinsville goblin} $<$/Description$>$
    \item $<$ UserId\textbf{$>$ DESKTOP-FBCFLB8\textbackslash kikoooioiooioi} $<$/UserId$>$
    \item $<$ RunLevel$>$ \textbf{HighestAvailable} $<$/RunLevel$>$
\end{itemize}

\begin{figure}[H]
    \centering
    \frame{\includegraphics[width=0.8\linewidth]{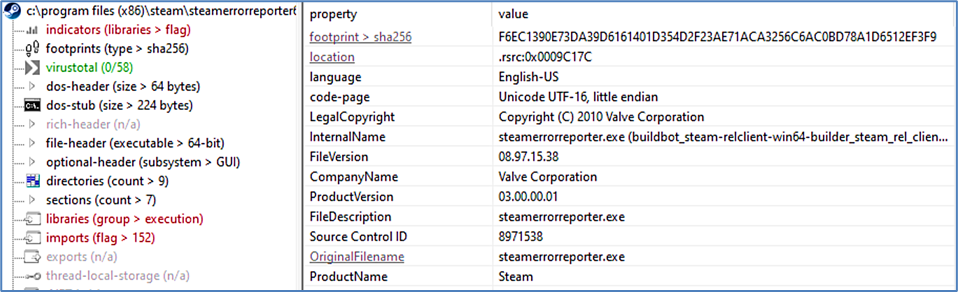}}
    \caption{\textit{steamerrorreporter.exe} version metadata.}
    \label{fig:sterr}
\end{figure}

\begin{figure}[H]
    \centering
    \frame{\includegraphics[width=0.8\linewidth]{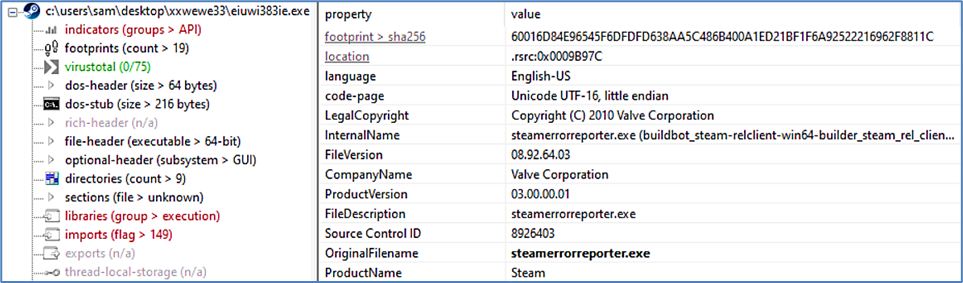}}
    \caption{\textit{eiuwi383ie.exe} version metadata.}
    \label{fig:sterrmal}
\end{figure}

Further investigations conducted on \textit{eiuwi383ie.exe} confirmed it as a legitimate instance of the well-known \textit{steamerrorreporter.exe} program, a tool used by \textit{Valve}’s \textit{Steam} \textit{marketplace} to manage execution errors. A side-by-side compilation comparison between the executable found in the malicious archive and the one deployed in the latest Steam installation revealed a key difference: the legitimate version is identified as \textit{08.97.15.28} (Fig. \ref{fig:sterr}), whereas the compromised version, dated \textit{August 3, 2024}, is labeled as \textit{08.92.64.03} (Fig. \ref{fig:sterrmal}).

Given the legitimate nature of this program, the malicious payload is likely embedded in one of the \textit{DLLs} included within the \textit{xxwewe33} archive. A further comparison of the imported functions between these files revealed that both versions interact with the same libraries, suggesting that the malicious modifications are concealed within the \textit{DLLs}, designed to blend in with the legitimate code.

\begin{figure}[H]
    \centering
    \frame{\includegraphics[width=0.9\linewidth]{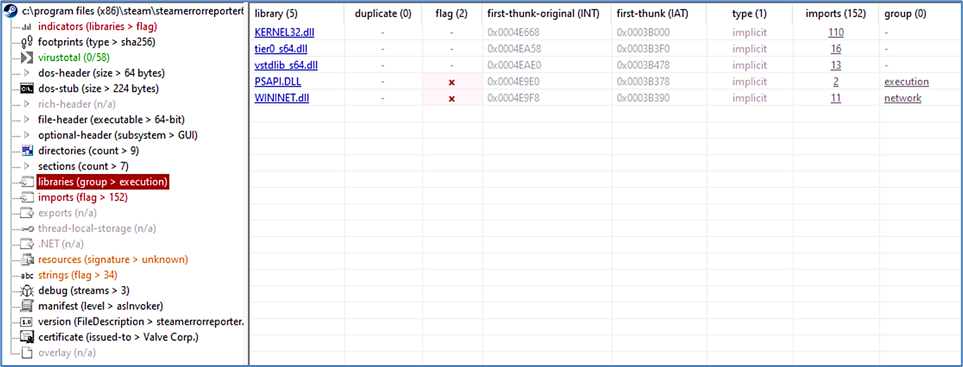}}
    \caption{Imports of legit \textit{steamerrorreporter.exe} which match the ones of the forged counterpart.}
    \label{fig:legsteam}
\end{figure}

\begin{figure}[H]
    \centering
    \frame{\includegraphics[width=1\linewidth]{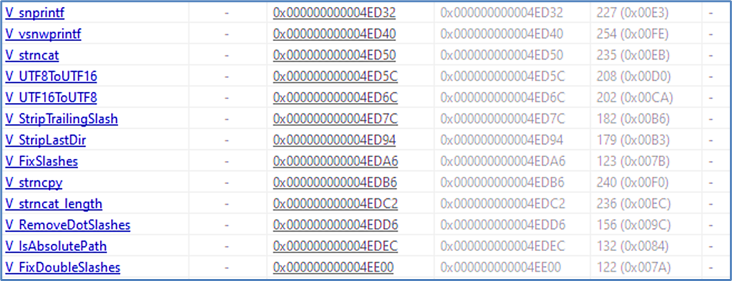}}
    \caption{\textit{steamerrorreporting.exe} importing functions from \textit{vstdlib\_s64.dll}.}
    \label{fig:legimports}
\end{figure}

\begin{figure}[H]
    \centering
    \frame{\includegraphics[width=1\linewidth]{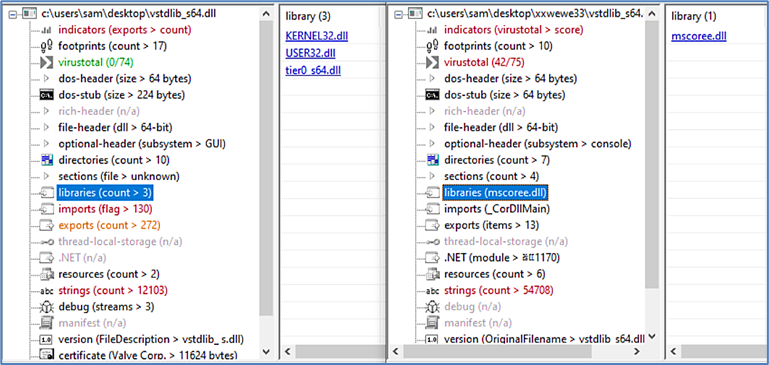}}
    \caption{Libraries count differences between legit and forged \textit{vstdlib\_s64.dll} instances.}
    \label{fig:malimports}
\end{figure}

\begin{figure}[H]
    \centering
    \frame{\includegraphics[width=0.7\linewidth]{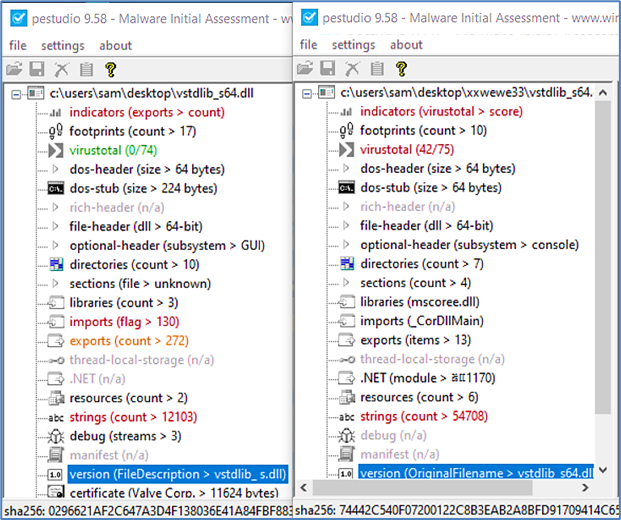}}
    \caption{Comprehensive side-by-side comparison between legit and forged \textit{vstdlib\_s64.dll}.}
    \label{fig:comp}
\end{figure}

Upon examining the elements depicted in Fig. \ref{fig:legsteam}, \textit{vstdlib\_s64.dll} was identified as the malicious file impersonating its legitimate counterpart. To ensure compatibility and avoid detection, the malicious version of this library exports only the specific functions required by \textit{steamerrorreporter.exe} (Fig. \ref{fig:legimports}), omitting any unnecessary methods (Fig. \ref{fig:malimports}). This selective export approach minimizes the likelihood of compatibility issues while maintaining the malware's stealth (fig. \ref{fig:comp}).

\begin{figure}[H]
    \centering
    \frame{\includegraphics[width=0.8\linewidth]{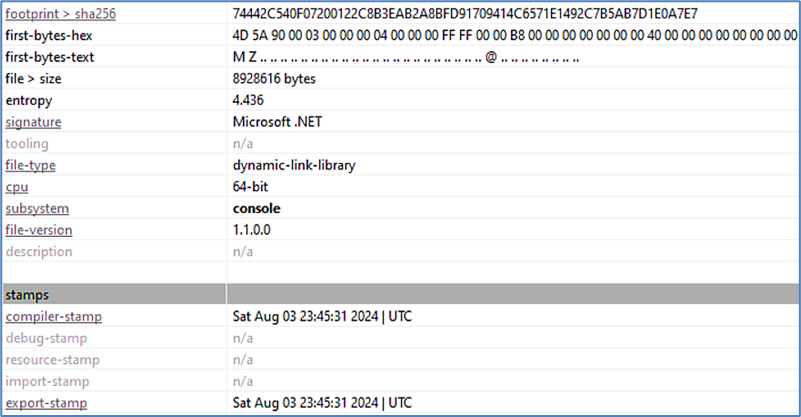}}
    \caption{Forged \textit{vstdlib\_s64.dll} compilation information.}
    \label{fig:forged}
\end{figure}

\begin{figure}[H]
    \centering
    \frame{\includegraphics[width=1\linewidth]{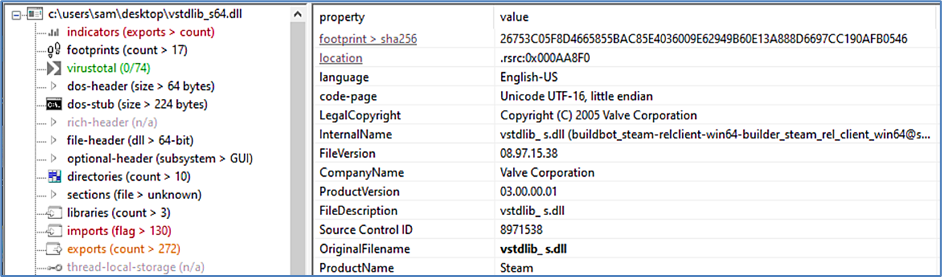}}
    \caption{Legit \textit{vstdlib\_s64.dll} details.}
    \label{fig:legdetails}
\end{figure}

\begin{figure}[H]
    \centering
    \frame{\includegraphics[width=1\linewidth]{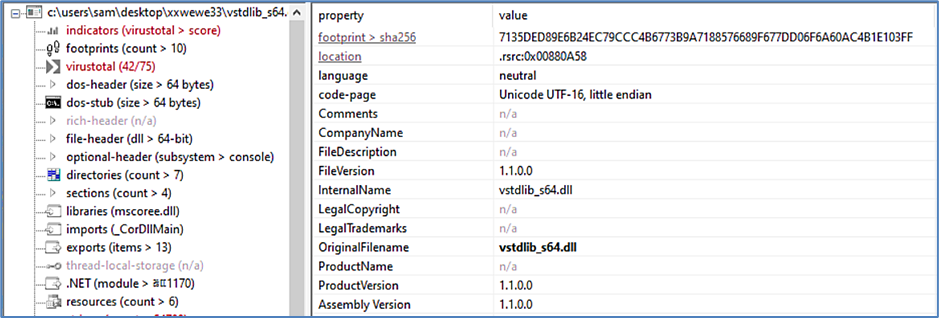}}
    \caption{Forged \textit{vstdlib\_s64.dll} details.}
    \label{fig:fordetails}
\end{figure}

The following additional main discrepancies were also identified (Fig. \ref{fig:forged}, Fig. \ref{fig:legdetails} and \ref{fig:fordetails}):
\begin{itemize}
    \item Imports and Exports sizes;
    \item Absence of \textit{.NET} information on the legit instance;
    \item Resources size;
    \item Version information;
    \item Digital Signature validity.
\end{itemize}

\begin{figure}[H]
    \centering
    \frame{\includegraphics[width=1\linewidth]{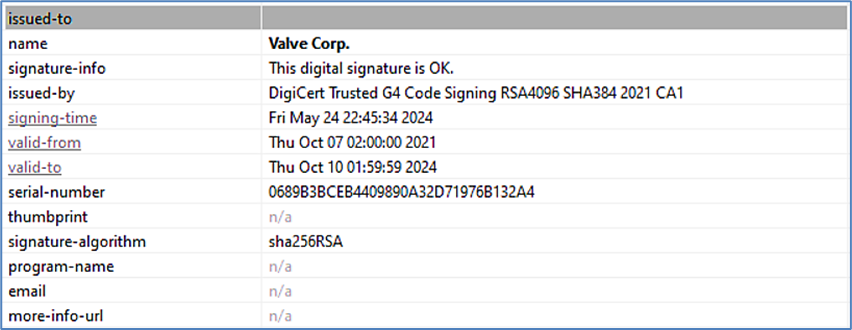}}
    \caption{Legit \textit{vstdlib\_s64.dll} Certificate information}
    \label{fig:legcert}
\end{figure}

\begin{figure}[H]
    \centering
    \frame{\includegraphics[width=1\linewidth]{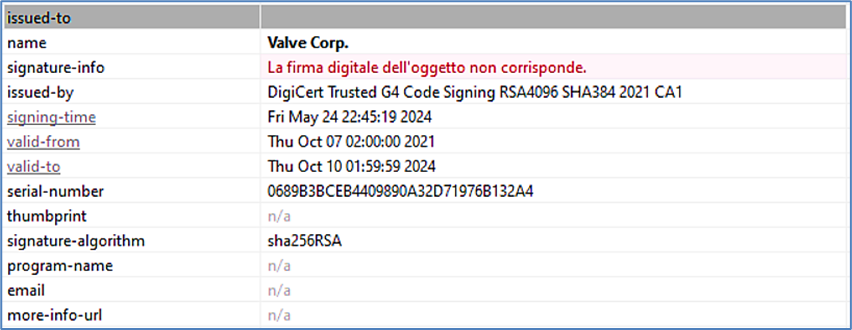}}
    \caption{Forged \textit{vstdlib\_s64.dll} Certificate information.}
    \label{fig:forcert}
\end{figure}

As noted earlier, the digital certificate associated with the fraudulent instance of \textit{vstdlib\_s64.dll} is invalid (Fig. \ref{fig:legcert} and Fig. \ref{fig:forcert}). Although both versions share the same signature, the hijacked version contains different code compared to its legitimate counterpart, preventing it from successfully authenticating the library. This indicates a poor security validation made by \textit{steamerrorreporter.exe}, which is not able to correctly manage forged signatures.

\begin{figure}[H]
     \centering
     \begin{subfigure}[b]{0.32\textwidth}
         \centering
         \frame{\includegraphics[width=\textwidth]{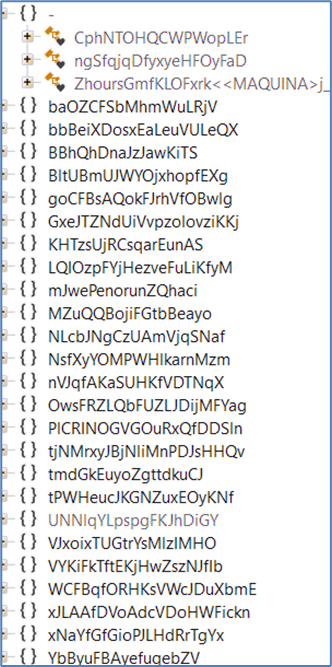}}
         \label{fig:loder}
     \end{subfigure}
     \hfill
     \begin{subfigure}[b]{0.35\textwidth}
         \centering
         \frame{\includegraphics[width=\textwidth]{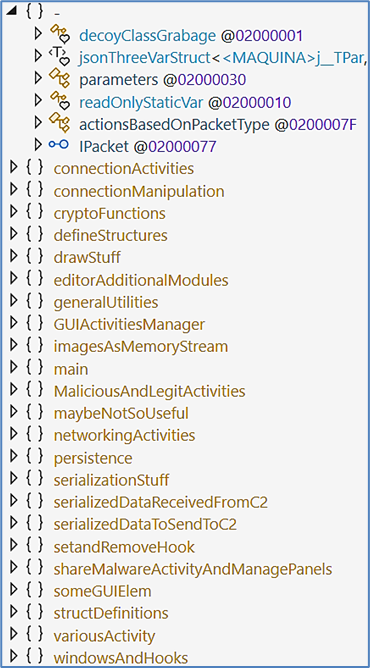}}
         \label{fig:pwrwx}
     \end{subfigure}
        \caption{Comparison of \textit{namespaces} before and after the refactoring of decompiled \textit{.NET} code.}
        \label{fig:two graphs}
\end{figure}

Since the targeted library has been compiled using a \textit{.NET} language, resulting in \textit{Intermediate Language} (\textit{IL}) code, it is feasible to reconstruct the original source code using well-known decompilers such as \textit{dnSpy}.
Fig. \ref{fig:two graphs} presents a side-by-side comparison of a freshly decompiled version of the library and the same file refactored after three days of analysis. By examining the names of the identified namespaces and classes, valuable insights into the threat’s capabilities and primary objectives on the victim’s workstation can be inferred (fIG. \ref{fig:two graphs}).

\begin{figure}[H]
    \centering
    \frame{\includegraphics[width=0.72\linewidth]{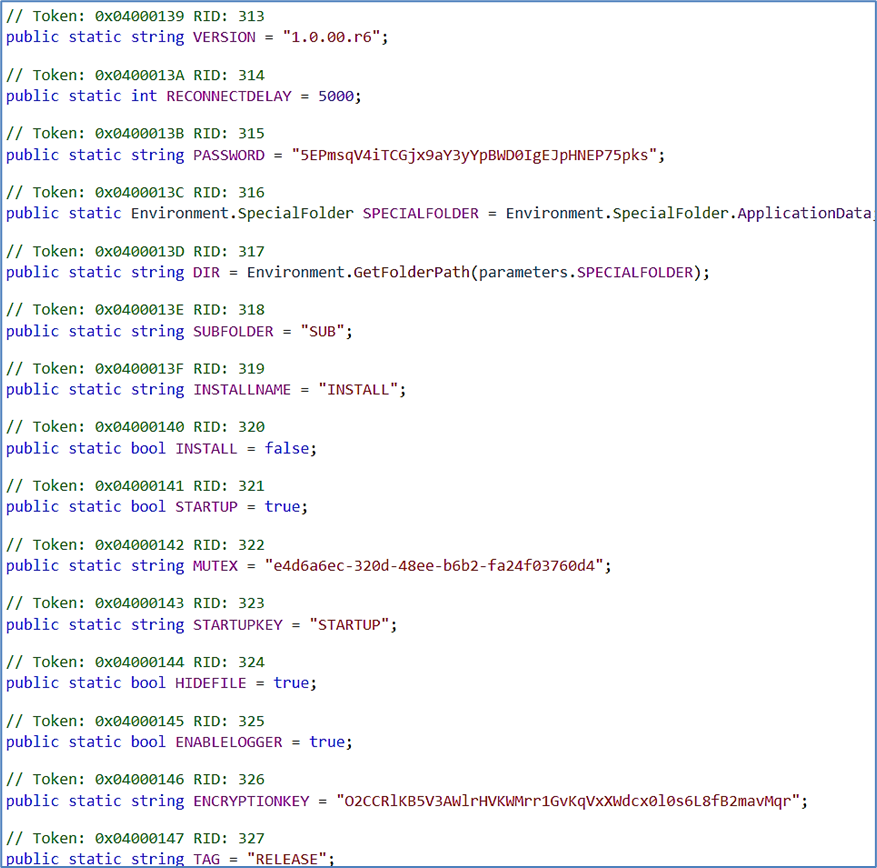}}
    \caption{Useful parameters set at the beginning of \textit{vstdlib\_s64.dll}.}
    \label{fig:var}
\end{figure}

The parameters class should be considered one of the initial elements for analysis to extract valuable information about this library. Within this class, several distinct strings are embedded, which may serve as potential \textit{Indicators of Compromise} (\textit{IoCs}) or variables relevant to reverse engineering. Notably, the most interesting strings include the following:
\begin{itemize}
    \item public static string VERSION = "1.0.00.r6";
    \item public static string PASSWORD = "5EPmsqV4iTCGjx9aY3yYpBWD0IgEJpHNEP\newline75pks";
    \item public static string INSTALLNAME = "INSTALL";
    \item public static string MUTEX = "e4d6a6ec-320d-48ee-b6b2-fa24f03760d4";
    \item public static string STARTUPKEY = "STARTUP";
    \item public static bool HIDEFILE = true;
    \item public static bool ENABLELOGGER = true;
    \item public static string ENCRYPTIONKEY = "O2CCRlKB5V3AWlrHVKWMrr1GvK\newline qVxXWdcx0l0s6L8fB2mavMqr";
\end{itemize}
Based on this evidence, it can be hypothesized that the threat establishes a remote channel with a reconnection delay of \textit{5000 ms}. Additionally, it utilizes the registry for persistence, writes a \textit{mutex} as a compromising signature into the system, hides itself as a concealed file, logs its operations, and encrypts data (Fig. \ref{fig:var}).

\begin{figure}[H]
    \centering
    \frame{\includegraphics[width=1\linewidth]{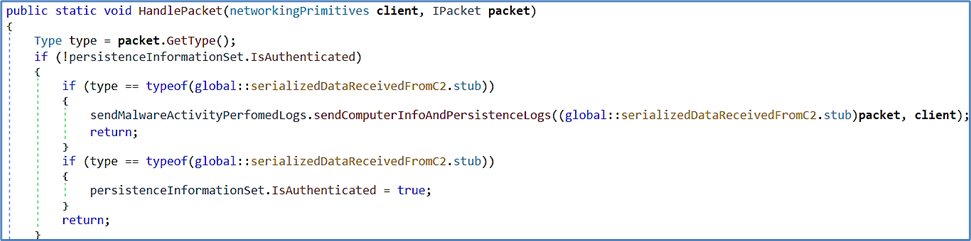}}
    \caption{Authentication mechanism based on exfiltration on basic asset's data.}
    \label{fig:auth}
\end{figure}

\begin{figure}[H]
    \centering
    \frame{\includegraphics[width=1\linewidth]{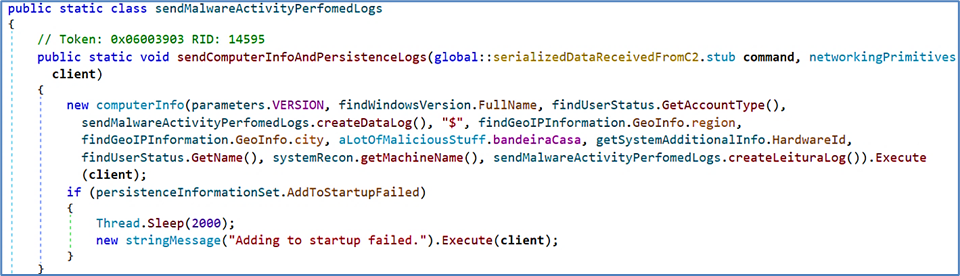}}
    \caption{Set of stolen information to be shared with the C\&C Server.}
    \label{fig:stoinfo}
\end{figure}

Within the same unnamed \textit{namespace}, another notable class for analysis is \textit{actionsBasedOnPacketType}. This enables the execution of various malicious activities depending on the type of serialized data received from the \textit{C\&C} Server. Initially, the code performs an authentication check (Fig. \ref{fig:auth}). This authentication process involves verifying whether a basic set of asset information has already been exfiltrated (Fig. \ref{fig:stoinfo}). If the code determines that authentication has not yet occurred, it will exfiltrate the necessary data and set the authentication variable to true.

\begin{figure}[H]
     \centering
     \begin{subfigure}[b]{0.32\textwidth}
         \centering
         \frame{\includegraphics[width=\textwidth]{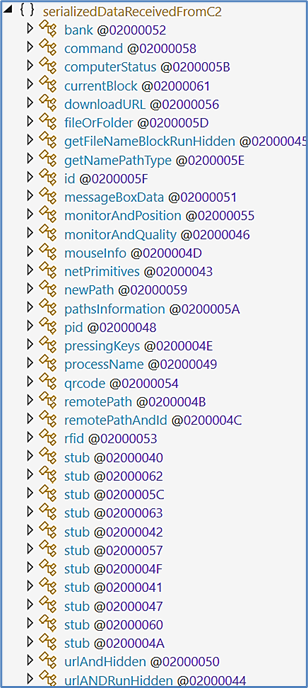}}
         \label{fig:lo}
     \end{subfigure}
     \hfill
     \begin{subfigure}[b]{0.4\textwidth}
         \centering
         \frame{\includegraphics[width=\textwidth]{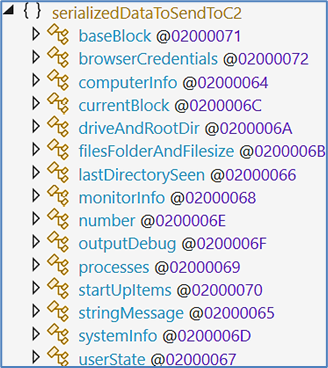}}
         \label{fig:pw}
     \end{subfigure}
        \caption{Serializable classes divided by their usage.}
        \label{fig:two graphs 2}
\end{figure}

\begin{figure}[H]
    \centering
    \frame{\includegraphics[width=0.8\linewidth]{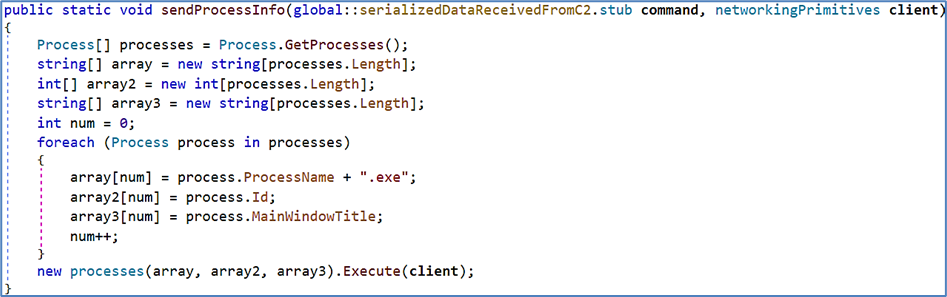}}
    \caption{An example of the structure of a stub serializable class.}
    \label{fig:stub}
\end{figure}

Given that this method was entirely focused on serialized classes, a thorough examination of these was conducted. This investigation revealed that the code defines two distinct \textit{namespaces} containing serializable classes (Fig. \ref{fig:two graphs 2}). The first panel illustrates all the serialized data that the threat receives from the C2 Server, providing insights into the class definitions. Among these, some classes are labeled as stub, a designation used for serialized classes that offer no meaningful information (Fig. \ref{fig:stub}). Conversely, the remaining serialized classes are responsible for exfiltrating information to the Command \& Control Server.

\begin{figure}[H]
    \centering
    \frame{\includegraphics[width=0.8\linewidth]{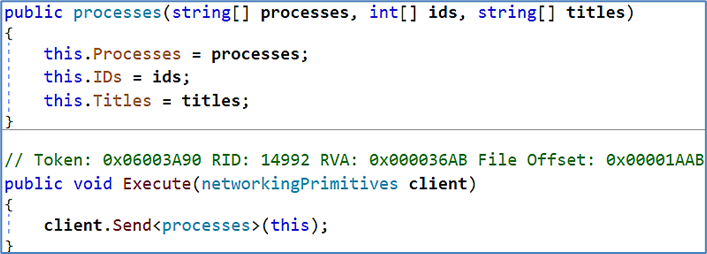}}
    \caption{Serializable class processes.}
    \label{fig:serial}
\end{figure}

Despite being organized under two different namespaces, all the classes share a common method named \textit{Execute}. This similarity is evident when comparing Fig. \ref{fig:stub}, which depicts a class from \textit{serializedDataReceivedFromC2} (Fig. \ref{fig:serial}), which shows a class from \textit{serializedDataToSendToC2}. Notably, within the provided code, references to the Execute method were found only in the context of the second \textit{namespace}, suggesting that its use may be specific to this \textit{namespace}. This design choice grants the \textit{Threat Actor} increased flexibility in refactoring the code, as it allows for the movement of different serializable classes between these namespaces without altering the fundamental method structure.

\begin{figure}[H]
    \centering
    \frame{\includegraphics[width=0.8\linewidth]{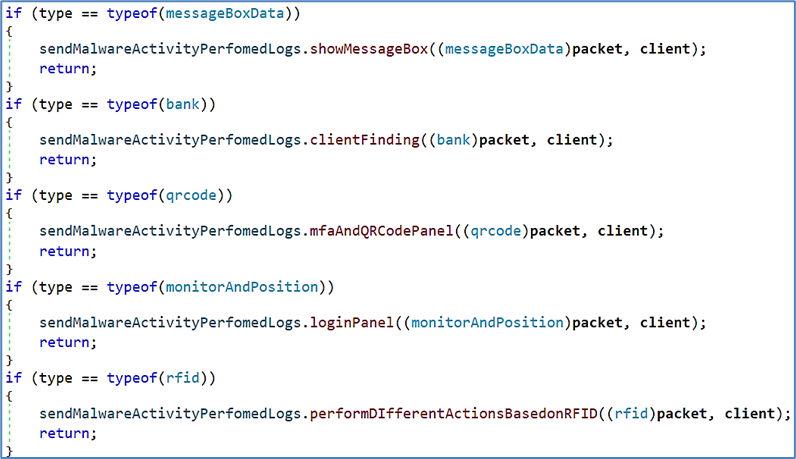}}
    \caption{Snippet of the function \textit{HandlePacket} of \textit{actionsBasedOnPacketType} class.}
    \label{fig:handle}
\end{figure}

\begin{figure}[H]
    \centering
    \frame{\includegraphics[width=0.9\linewidth]{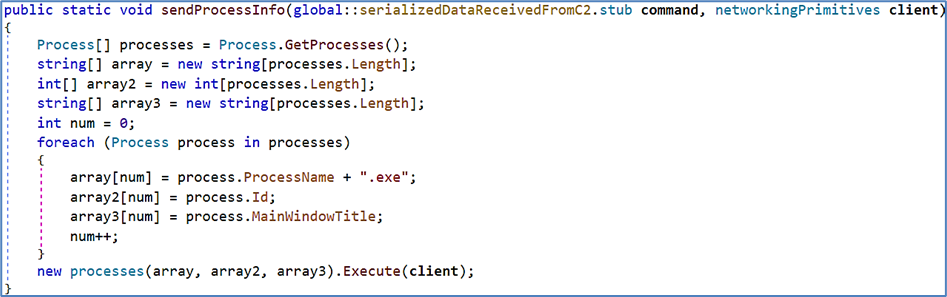}}
    \caption{Function used to gather Processes information.}
    \label{fig:23}
\end{figure}

\begin{figure}[H]
    \centering
    \frame{\includegraphics[width=0.9\linewidth]{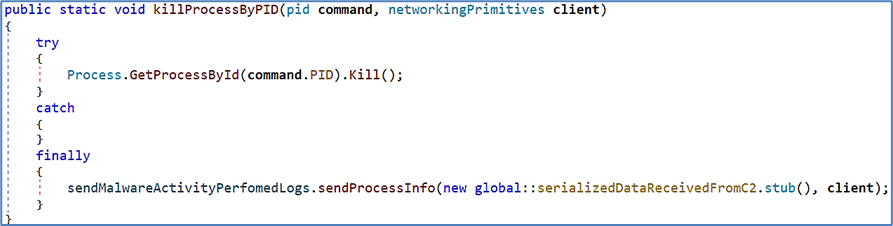}}
    \caption{Function used to kill a process given its \textit{PID}.}
    \label{fig:24}
\end{figure}

\begin{figure}[H]
    \centering
    \frame{\includegraphics[width=1\linewidth]{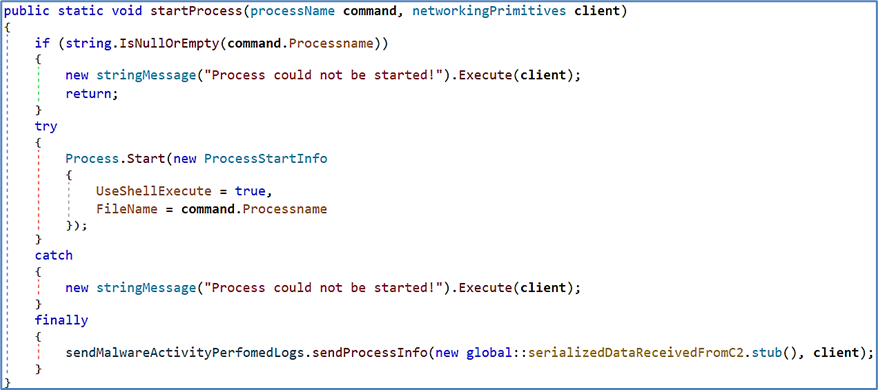}}
    \caption{Function used to start a new process}
    \label{fig:25}
\end{figure}

\begin{figure}[H]
    \centering
    \frame{\includegraphics[width=1\linewidth]{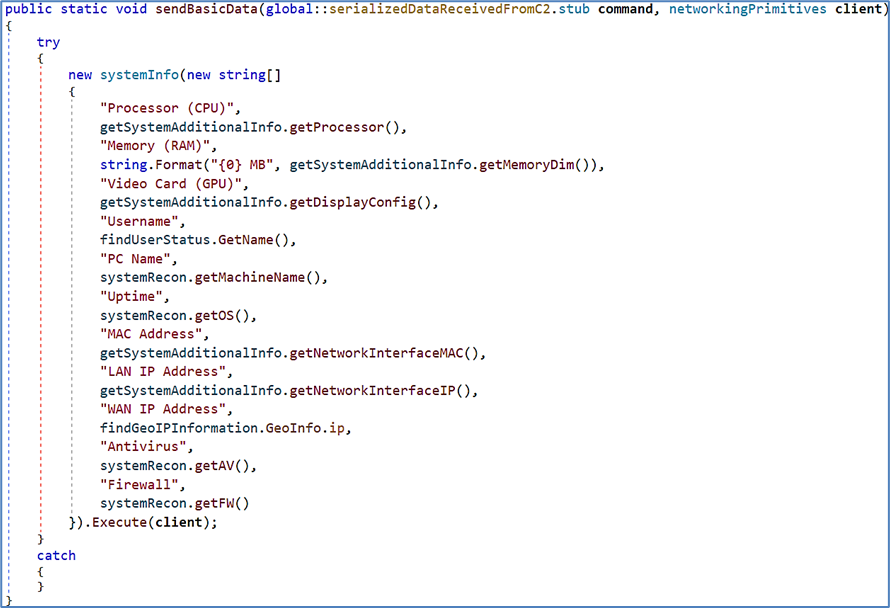}}
    \caption{Additional basic information stolen and sent to \textit{C\&C}.}
    \label{fig:26}
\end{figure}

\begin{figure}[H]
    \centering
    \frame{\includegraphics[width=0.7\linewidth]{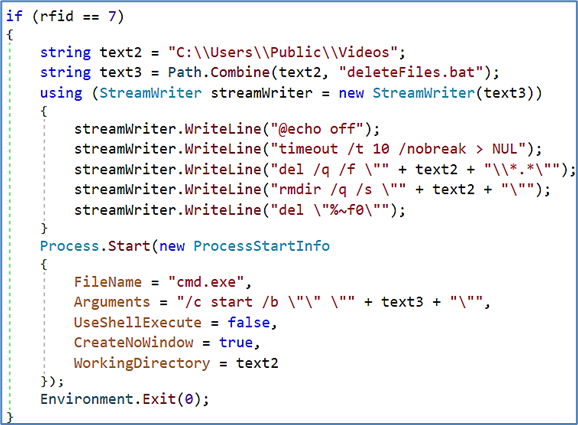}}
    \caption{\textit{RFID} value 7 creates a \textit{deleteFiles.bat} file to delete Videos folder.}
    \label{fig:27}
\end{figure}

\begin{figure}[H]
    \centering
    \frame{\includegraphics[width=0.7\linewidth]{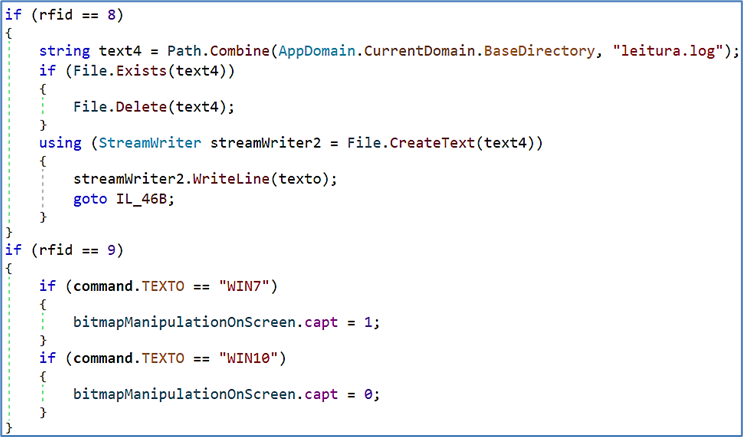}}
    \caption{\textit{RFID} values 8 and 9 allow to write log files and vary its behavior.}
    \label{fig:28}
\end{figure}

\begin{figure}[H]
    \centering
    \frame{\includegraphics[width=0.8\linewidth]{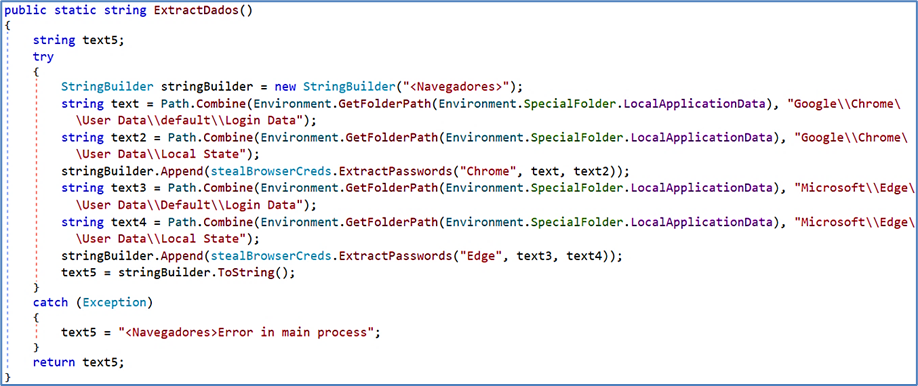}}
    \caption{\textit{RFID} value 11, information on installed browsers and stored credentials.}
    \label{fig:29}
\end{figure}

\begin{figure}[H]
    \centering
    \frame{\includegraphics[width=0.8\linewidth]{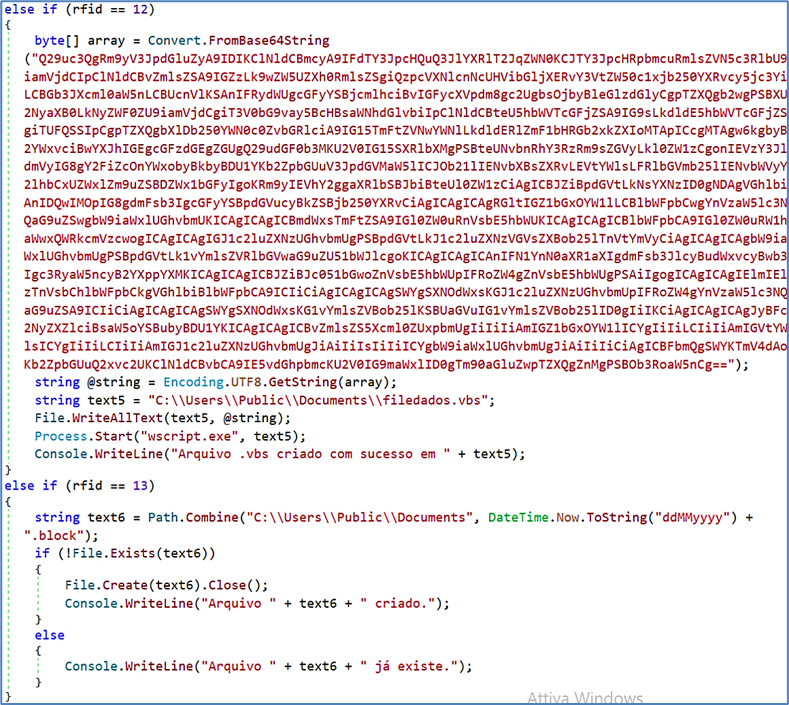}}
    \caption{\textit{RFID} value 12 allows to decode a \textit{Base64} text into \textit{filedados.vbs} which aims to steal outlook contacts.}
    \label{fig:30}
\end{figure}

\begin{figure}[H]
    \centering
    \frame{\includegraphics[width=1\linewidth]{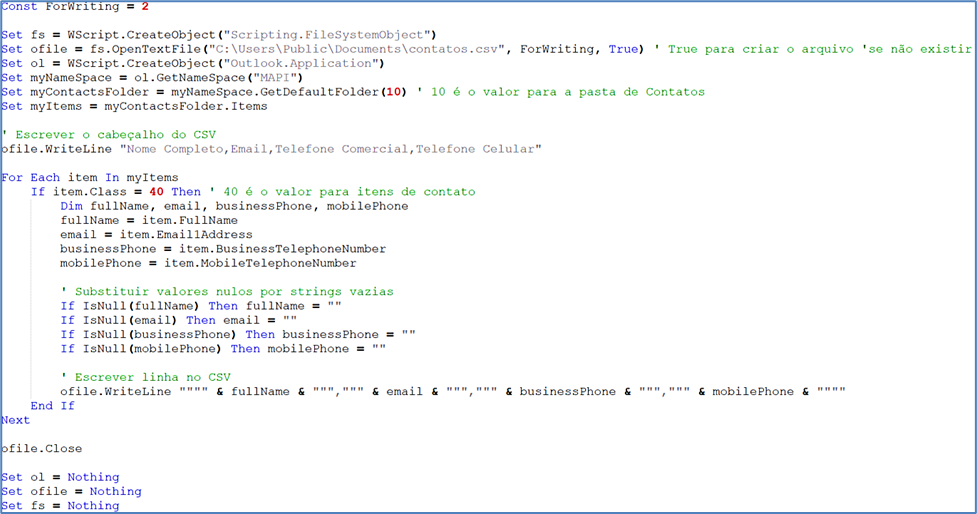}}
    \caption{Decoded \textit{Base64} content of \textit{filedados.vbs}.}
    \label{fig:31}
\end{figure}

\begin{figure}[H]
    \centering
    \frame{\includegraphics[width=0.8\linewidth]{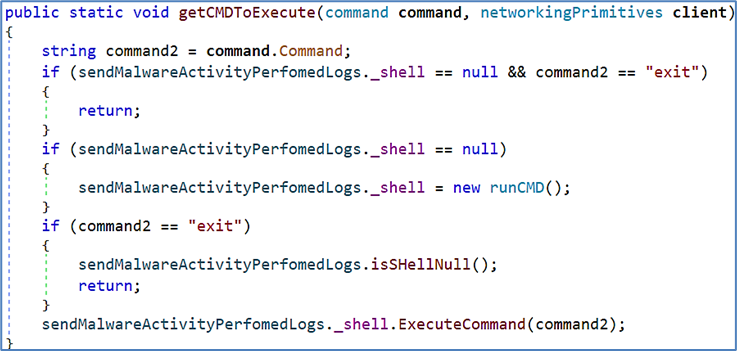}}
    \caption{Function used to run \textit{cmd} commands provided by the \textit{C\&C} Server.}
    \label{fig:32}
\end{figure}

\begin{figure}[H]
    \centering
    \frame{\includegraphics[width=0.8\linewidth]{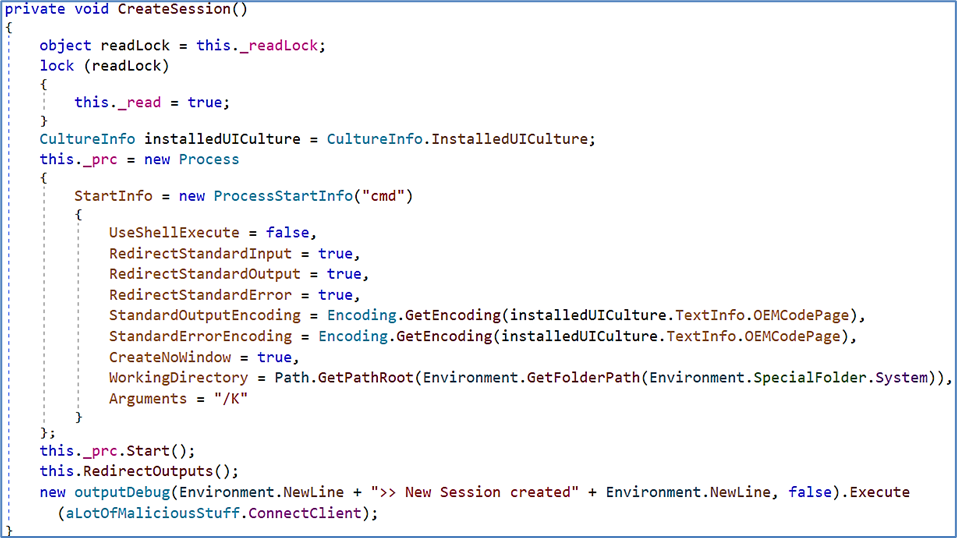}}
    \caption{Primitive used as building block to run \textit{cmd} commands.}
    \label{fig:33}
\end{figure}

As briefly discussed previously (Fig. \ref{fig:handle}), \textit{HandlePacket} method can perform a wide set of different actions, just as:
\begin{itemize}
    \item Reconnect or do not reconnect to server;
    \item Uninstall from persistence while getting, setting and deleting elements from startup registry entries;
    \item Exfil monitor settings;
    \item Exfil processes (Fig. \ref{fig:23});
    \item Exfil external drive info;
    \item Kill process by PID (Fig. \ref{fig:24});
    \item Start a process (Fig. \ref{fig:25});
    \item Exfil additional asset information (Fig. \ref{fig:26});
    \item Interact with input devices;
    \item Show Message Box with custom text;
    \item Show MFA Form;
    \item Show authentication QR Codes forms;
    \item Basing on an additional embedded parameter, named RFID, additional malicious activities can be performed:
    \begin{itemize}
        \item Deleting Video folder (Fig. \ref{fig:27});
        \item Save log files (Fig. \ref{fig:28});
        \item Adapt behavior to different Windows versions (Fig. \ref{fig:28});
        \item Exfil information on installed browsers and stored credentials (Fig. \ref{fig:29});
        \item Steal Outlook contacts and exfiltrate them (Fig. \ref{fig:30} and Fig. \ref{fig:31}).
        \item Keylogging
    \end{itemize}
    \item Execute shell commands (Fig. \ref{fig:32} and Fig. \ref{fig:33});
    \item Rename files and directories;
    \item Shutdown and Reboot the system;
    \item Cancel downloads;
    \item Get, Add and Remove AutoStart items;
    \item Exfil Log files, browsers’ and bank credentials;
    \item Connect to Reverse proxies.
\end{itemize}

\begin{figure}[H]
    \centering
    \frame{\includegraphics[width=1\linewidth]{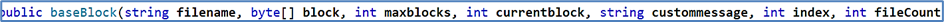}}
    \caption{\textit{baseBlock} structure}
    \label{fig:35}
\end{figure}

\begin{figure}[H]
    \centering
    \frame{\includegraphics[width=0.9\linewidth]{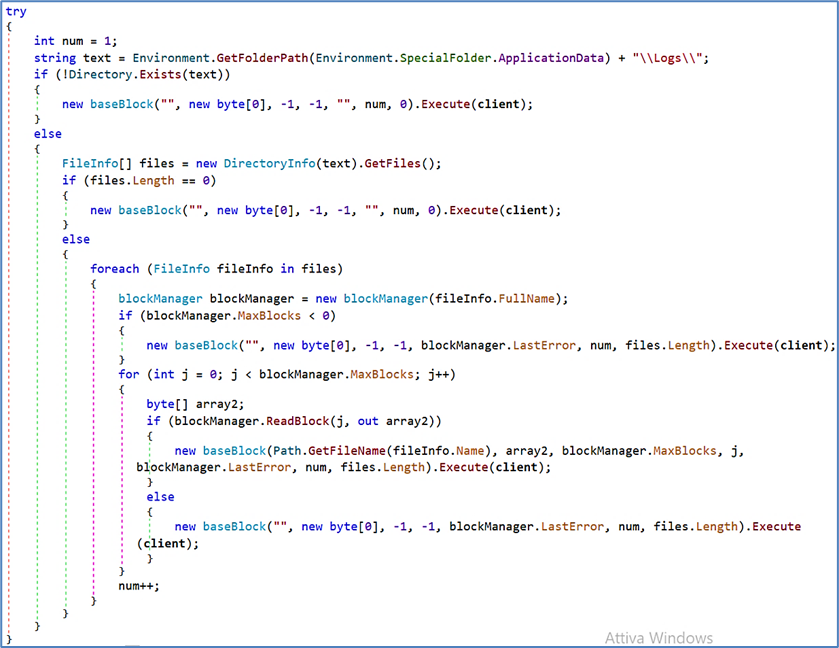}}
    \caption{Analysis of the function which creates and exfiltrates logs as blocks.}
    \label{fig:34}
\end{figure}

\begin{figure}[H]
    \centering
    \frame{\includegraphics[width=0.65\linewidth]{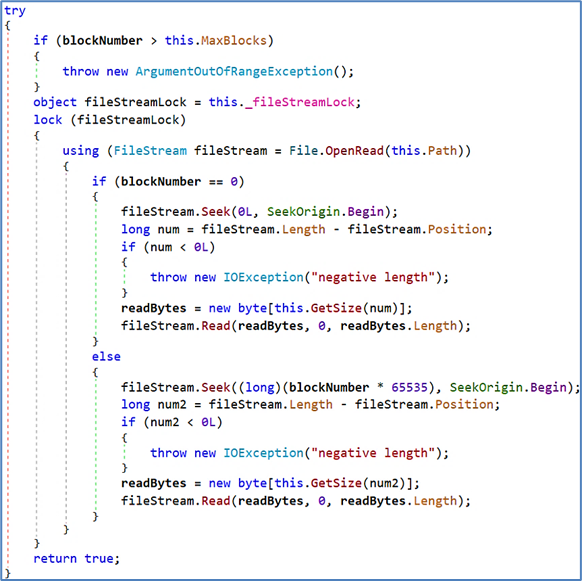}}
    \caption{How a \textit{baseManager} reads a block.}
    \label{fig:36}
\end{figure}

In certain instances, rather than directly handling the data, this threat manipulates it by segmenting it into blocks (Fig. \ref{fig:35}, Fig. \ref{fig:34} and Fig. \ref{fig:36}), each with a maximum size of \textit{65'535} bits. This approach is likely employed to evade network detection. For example, when exfiltrating logs (as illustrated in Figure 34), the threat constructs a corresponding \textit{BlockManager} for each file within the Logs folder. This method of data management aids in obfuscating the exfiltration process and avoiding detection by distributing the data across multiple smaller segments. 
The \textit{BlockManager} functions as a supervisor, overseeing the segmentation and proper mapping of each data block. For each segment into which a file is divided, an instance of the \textit{baseBlock} class is created and initialized with detailed information about the block. This includes the filename, the block's byte data, its index, any operational errors encountered, and the overall length of the file.

\begin{figure}[H]
    \centering
    \frame{\includegraphics[width=0.8\linewidth]{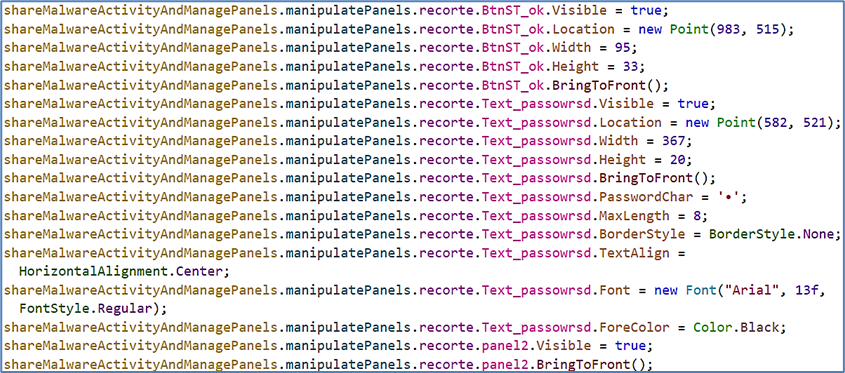}}
    \caption{Snippet of a function used to display login forms.}
    \label{fig:login}
\end{figure}

All the mechanisms outlined in the previous section are associated with activities aimed at stealing banking credentials from victims in \textit{LATAM}. Additionally, the extensive use of \textit{Portuguese}/\textit{Brazilian} comments and strings within the code suggests that the Threat Actor behind this malicious \textit{DLL} may originate from \textit{Brazil}. However, this remains a hypothesis based on the available evidence.  Further analysis then revealed a series of \textit{Windows} that are loaded to mimic login forms for various banks (Fig. \ref{fig:login}).

\begin{figure}[H]
    \centering
    \frame{\includegraphics[width=0.8\linewidth]{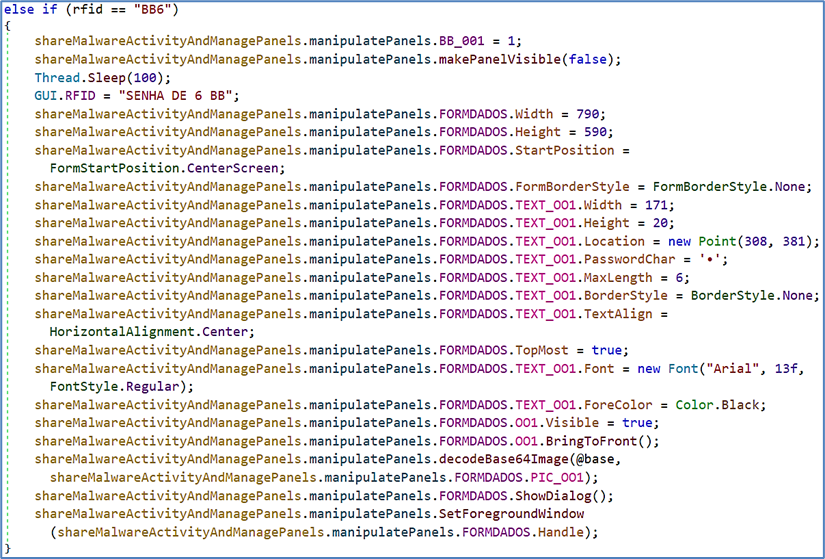}}
    \caption{Snippet of code used to show a fraudulent 6-digit \textit{MFA} prompt for \textit{Banco do Brazil} login.}
    \label{fig:6dig}
\end{figure}

\begin{figure}[H]
    \centering
    \frame{\includegraphics[width=0.8\linewidth]{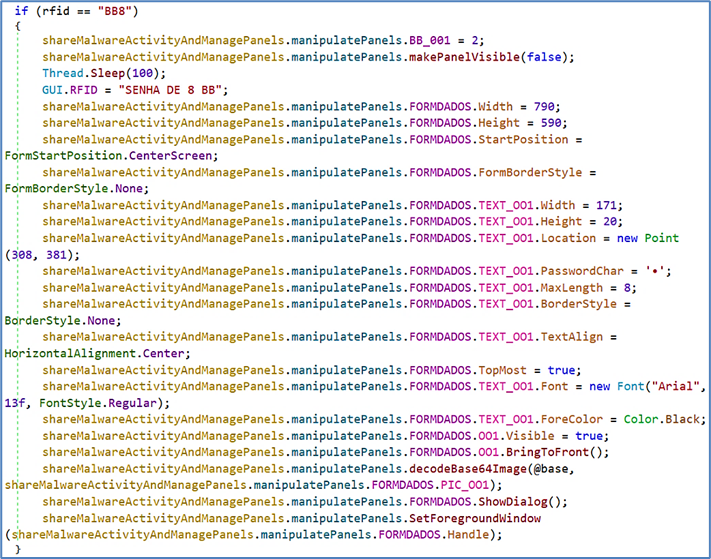}}
    \caption{Snippet of code used to show a fraudulent 8-digit \textit{MFA} prompt for \textit{Banco do Brazil} login}
    \label{fig:8dig}
\end{figure}

After entering their credentials, victims are confronted with a multi-factor authentication (MFA) window tailored specifically to their bank account. The code for each bank features slightly different implementations to reflect the variations in MFA processes. For instance, in cases \textit{BB6} (Fig. \ref{fig:6dig}) and \textit{BB8} (Fig. \ref{fig:8dig}), the only visible difference is in the \textit{MaxLength} parameter of the \textit{TEXT\_001} input field, which is set to 6 and 8 characters, respectively. 

This indicates that the Threat Actor did not invest significant effort into producing efficient or elegant source code. Additionally, this lack of refinement is not an isolated issue; throughout the analysis, numerous examples of inefficient and inelegant functions, along with inadequate error handling and the use of deprecated methods, were identified.

\begin{figure}[H]
     \centering
     \begin{subfigure}[b]{0.2\textwidth}
         \centering
         \frame{\includegraphics[width=\textwidth]{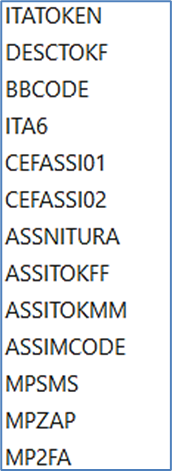}}
         \label{fig:loader}
     \end{subfigure}
     \hfill
     \begin{subfigure}[b]{0.21\textwidth}
         \centering
         \frame{\includegraphics[width=\textwidth]{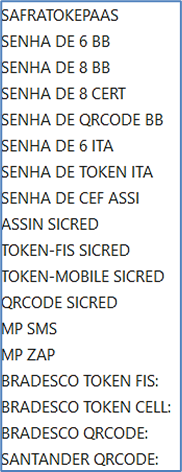}}
         \label{fig:pwshrwx}
     \end{subfigure}
        \caption{List of codes used to distinguish between banks and their \textit{MFA} prompts.}
        \label{fig:MFAs}
\end{figure}

As illustrated in Fig. \ref{fig:MFAs}, the extracted code used by the \textit{QRCode} function to display \textit{MFA} prompts reveals potential impacts on several financial institutions:
\begin{itemize}
    \item \textit{Banco do Brasil SA}
    \item \textit{Itaú Unibanco Holding SA}
    \item \textit{Banco Bradesco SA}
    \item \textit{Caixa Econômica Federal}
    \item \textit{Banco Santander Chile/Mexico}
    \item \textit{Banco Safra}
    \item \textit{Banco Cooperativo Sicredi}
\end{itemize}

\begin{figure}[H]
    \centering
    \frame{\includegraphics[width=0.8\linewidth]{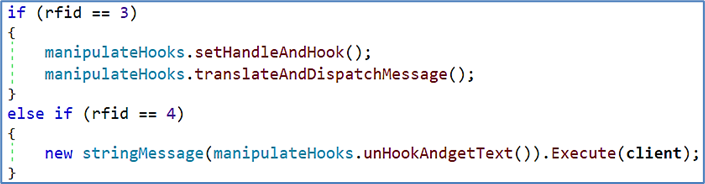}}
    \caption{\textit{RFID} 3 and 4 are used to code the keylogging activation and acquired data exfiltration.}
    \label{fig:keylgg}
\end{figure}

\begin{figure}[H]
    \centering
    \frame{\includegraphics[width=0.9\linewidth]{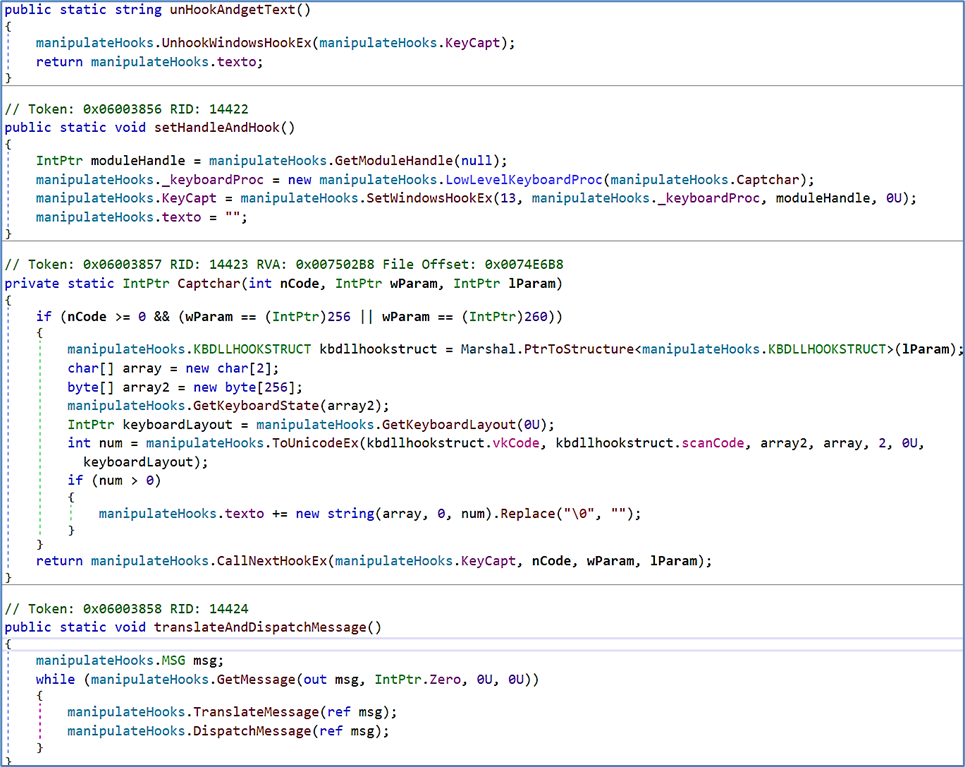}}
    \caption{Main keylogging methods defined in the subjected malicious \textit{DLL}.}
    \label{fig:keydef}
\end{figure}

Another notable capability of this \textit{DLL} pertains to its keylogging functionality (Fig. \ref{fig:keylgg} and Fig. \ref{fig:keydef}). The \textit{Command \& Control} server utilizes an \textit{RFID} code of 3 to enable the keylogging feature. Conversely, an \textit{RFID} code of 4 triggers the \textit{DLL} to transmit the captured text back to the server. The \textit{DLL} leverages the \textit{translateAndDispatchMessage} method, which calls functions from \textit{user32.dll}, including \textit{GetMessage}, \textit{TranslateMessage}, and \textit{DispatchMessage}. This method is used within a broader context involving hooks and input manipulation, suggesting that this loop may also handle messages related to these hooks, such as custom keyboard or mouse events.
In addition to the previously identified capabilities, it is important to note that this malicious DLL also can manipulate and gather screen information. It can alter the mouse position and configurations, further expanding its potential for malicious activities.

\begin{figure}[H]
    \centering
    \frame{\includegraphics[width=0.75\linewidth]{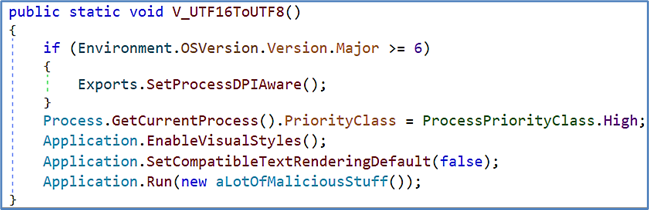}}
    \caption{\textit{V\_UTF16ToUTF8} exported function.}
    \label{fig:dllimpl}
\end{figure}

After an initial understanding of the threat’s capabilities, it was needed to find out how it really operates when executed. To do so, it was needed to find the \textit{namespace} exporting the functions requested by \textit{steamerrorreporting.exe}. Once found, it was possible to find how none but one of them was really implemented (Fig. \ref{fig:dllimpl}).

\begin{figure}[H]
    \centering
    \frame{\includegraphics[width=0.7\linewidth]{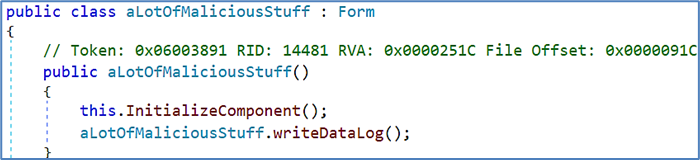}}
    \caption{\textit{aLotOfMaliciousStuff} initialization.}
    \label{fig:malinit}
\end{figure}

Once the malicious \textit{DLL} is loaded, it initializes a class designated as \textit{aLotOfMaliciousStuff}, which constructs an interactive form for the victim (Fig. \ref{fig:malinit}).

\begin{figure}
    \centering
    \frame{\includegraphics[width=0.8\linewidth]{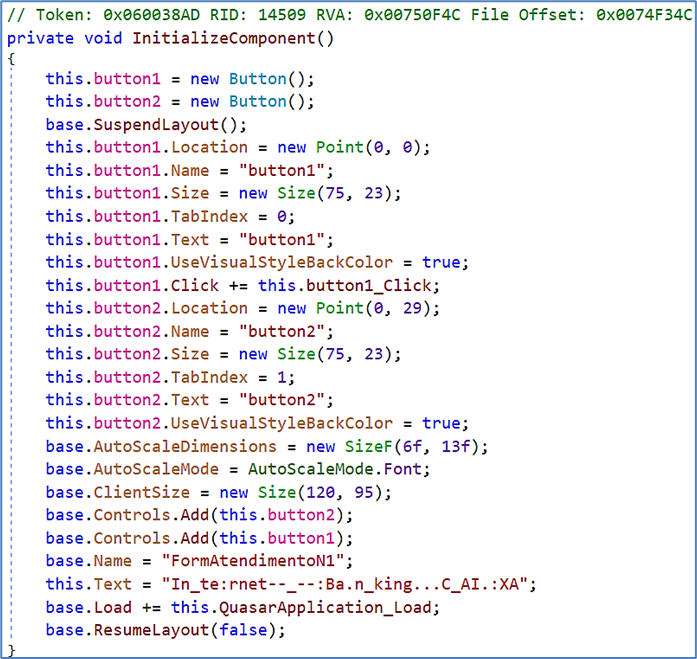}}
    \caption{First form loaded by forged \textit{vstdlib\_s64.dll}.}
    \label{fig:forgeddll}
\end{figure}

The function that initializes this form shows an interesting string mentioning \textit{Internet Banking CAIXA}, used to mimic the Spanish multinational financial service company (Fig. \ref{fig:forgeddll})\footnote{https://www.caixa.gov.br/}.

\begin{figure}[H]
    \centering
    \frame{\includegraphics[width=0.9\linewidth]{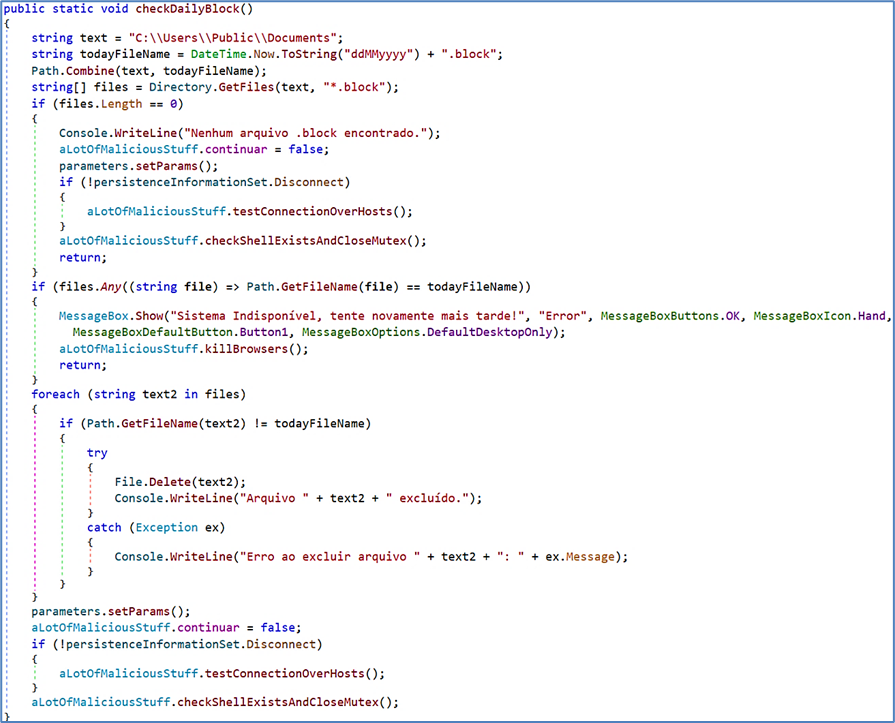}}
    \caption{Function used to check for the existence of a \textit{daily block}.}
    \label{fig:chkdly}
\end{figure}

\begin{figure}[H]
    \centering
    \frame{\includegraphics[width=0.9\linewidth]{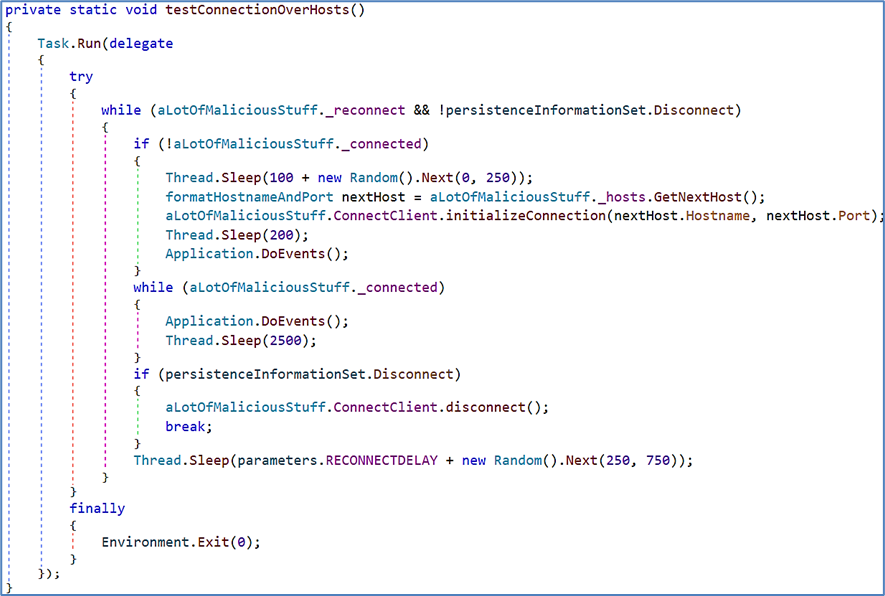}}
    \caption{\textit{testConnectionOverHosts} function}
    \label{fig:tstconn}
\end{figure}

\begin{figure}[H]
    \centering
    \frame{\includegraphics[width=0.6\linewidth]{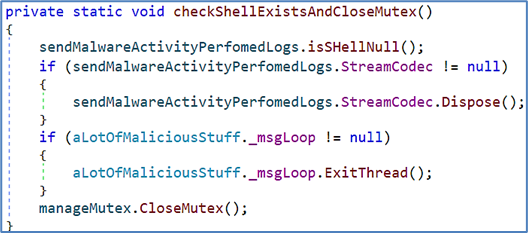}}
    \caption{\textit{checkShellExistAndCloseMutex} function}
    \label{fig:chkshell}
\end{figure}

Once the victim clicks on \textit{Button\_1}, a routine named \textit{checkDailyBlock} is executed. This routine searches for the presence of a \textit{daily .block} file in the Documents folder (Fig. \ref{fig:chkdly}). If the file is not found, the function terminates. Before exiting, it attempts to reconnect to its \textit{Command \& Control} server if the connection had been previously lost. The routine then assigns the parameters indicated in Fig. \ref{fig:var} to class variables. Should the connection to the \textit{C2} server have been interrupted, it makes an additional attempt to re-establish it (Fig. \ref{fig:tstconn}) and verifies the existence of a handle to a shell instance (Fig. \ref{fig:chkshell}). 

\begin{figure}[H]
    \centering
    \frame{\includegraphics[width=0.7\linewidth]{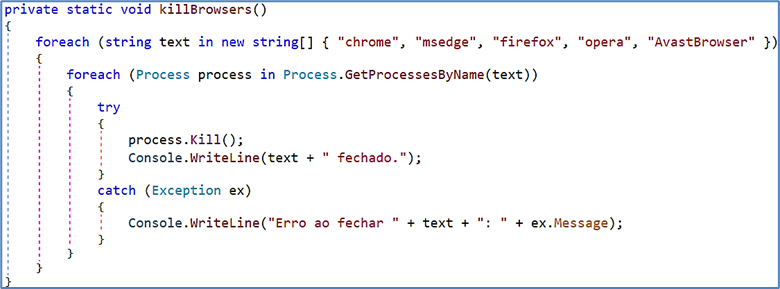}}
    \caption{Function used to kill any active instance of the hardcoded browsers.}
    \label{fig:browser}
\end{figure}

If the \textit{daily .block} file is present, the routine displays a message box indicating an error, advising the user to try again later. Following this, it attempts to terminate any running instances of \textit{Chrome}, \textit{Edge}, \textit{Firefox}, \textit{Opera}, and \textit{Avast Browser} (Fig. \ref{fig:browser}). Before exiting, it performs the same actions as described earlier: setting parameters, verifying the connection status, and checking for the existence of a handle to a shell instance.

\begin{figure}[H]
    \centering
    \frame{\includegraphics[width=0.8\linewidth]{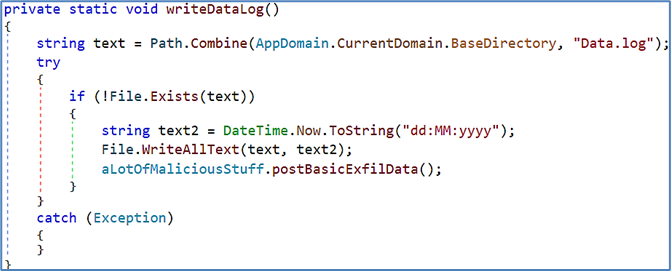}}
    \caption{Function used to write a log file containing the today’s date.}
    \label{fig:mutex}
\end{figure}

Upon invocation, the class also triggers a function named \textit{writeDataLog} (Fig. \ref{fig:mutex}). This function first checks for the presence of a Data.log file in the directory where the DLL was executed. If the file already exists, the function simply exits. However, if the file is not found, it creates the file, initializes it with the current date, and then calls another function to send basic information to the \textit{Command \& Control} (C2) server. The existence of this Data.log file serves as a \textit{mutex}, enabling the DLL to determine whether the \textit{C2} server has already been informed of the host's compromise. Additionally, since the file is not removed, it functions as a timestamp for the infection date.

\begin{figure}[H]
    \centering
    \frame{\includegraphics[width=0.9\linewidth]{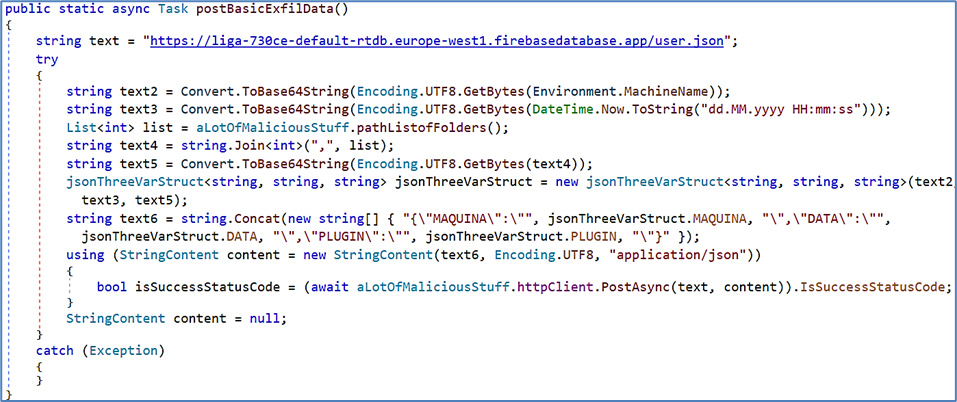}}
    \caption{Function used to exfiltrate data to \textit{C\&C}.}
    \label{fig:exfilcc}
\end{figure}

\textit{postBasicExfilData} gathers the following three main asset’s information:
\begin{itemize}
    \item Machine Name
    \item Actual Date and Time
    \item Presence of known banking software
\end{itemize}
All these information gets stored in a \textit{json} structure, where the content is \textit{Base64} encoded, and posted as an \textit{async} call to the \textit{C\&C} Server. The following are the respective used to map these values (Fig. \ref{fig:exfilcc}):
\begin{itemize}
    \item \textit{MAQUINA}
    \item \textit{DATA}
    \item \textit{PLUGIN}
\end{itemize}

\begin{figure}[H]
    \centering
    \frame{\includegraphics[width=0.8\linewidth]{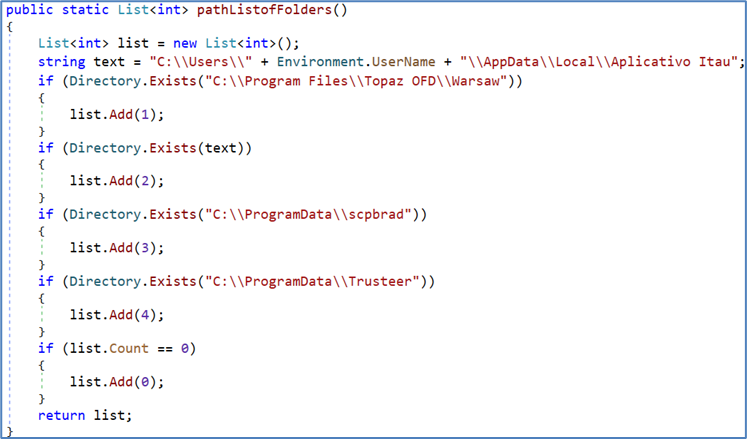}}
    \caption{Function used to map installed banking software.}
    \label{fig:banksoft}
\end{figure}

The following function is employed to verify whether some security-related software is installed on the compromised asset. All of them seems to be distributed by desktop home-banking solutions which require this security plugins to be installed (Fig. \ref{fig:banksoft}). When one of them is found, a special code gets added to the list that is passed to the callee function and sent to the \textit{C2} Server:
\begin{itemize}
    \item \textit{Aplicativo Itau}: Home-banking software related to \textit{Itaú Unibanco Holding SA};
    \item \textit{Topaz OFD}: Banking security module which some banks, e.g. \textit{Banco do Brazil}, require to be installed;
    \item \textit{Trusteer}: Security software, developed by \textit{IBM}, which is installed by some banks to enhance user’s security over frauds and threat infections;
    \item \textit{scpbrad}: Associated to the \textit{ScpSecurityService} software which is developed and distributed by \textit{Banco Bradesco S.A.}
\end{itemize}

\begin{figure}[H]
    \centering
    \frame{\includegraphics[width=0.8\linewidth]{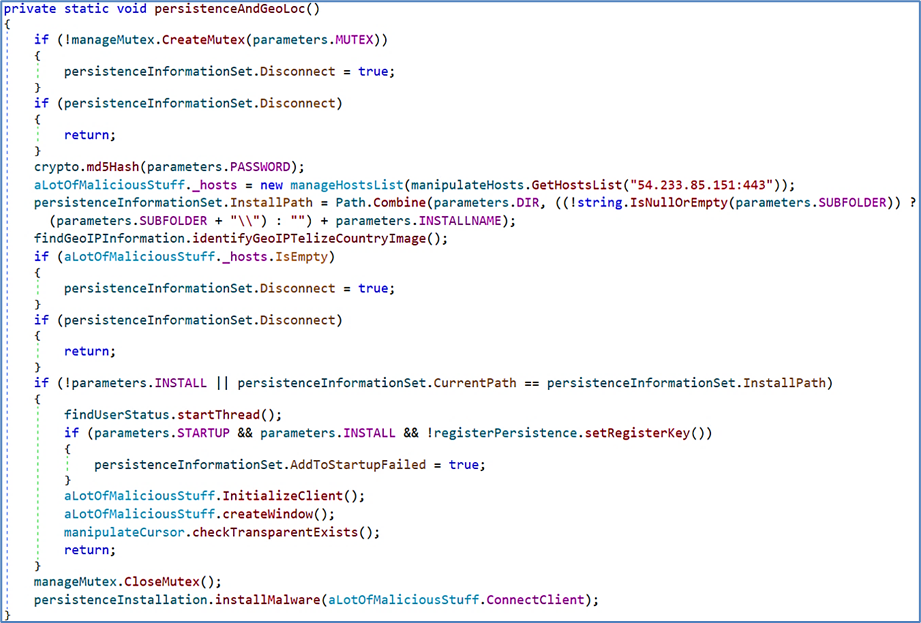}}
    \caption{Function used to localize the victim and achieve persistence into the system.}
    \label{fig:pers}
\end{figure}

Meanwhile, the \textit{OnLoad} function undertakes tasks related to the geographic identification of the victim’s IP address and the implementation of persistence mechanisms (Fig. \ref{fig:pers}).

\begin{figure}[H]
    \centering
    \frame{\includegraphics[width=0.8\linewidth]{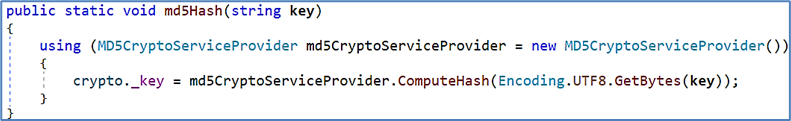}}
    \caption{\textit{MD5} hash function to manipulate the hardcoded \textit{PASSWORD} parameter.}
    \label{fig:password}
\end{figure}

Initially, the function attempts to create a \textit{mutex} using the name specified in the initial configuration class. If the \textit{mutex} creation fails, the function disconnects from the \textit{C2} server and terminates. However, if the \textit{mutex} is successfully created, the process proceeds by generating a new cryptographic key through the \textit{MD5} hashing of the embedded \textit{PASSWORD} parameter (Fig. \ref{fig:password}). Subsequently, it updates the host list with the \textit{C\&C} IP address and port.

\textit{C:\textbackslash Users\textbackslash $<$username$>$\textbackslash AppData\textbackslash Roaming\textbackslash SUB\textbackslash INSTALL} is chosen as value for the \textit{InstallPath} variable. 

\begin{figure}[H]
    \centering
    \frame{\includegraphics[width=1\linewidth]{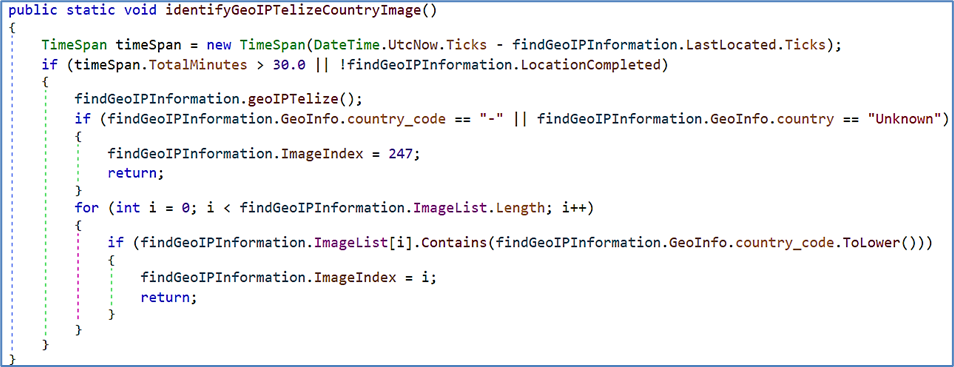}}
    \caption{Function to identify victim’s location and find an image for its country’s flag.}
    \label{fig:flag}
\end{figure}

\begin{figure}[H]
    \centering
    \frame{\includegraphics[width=0.9\linewidth]{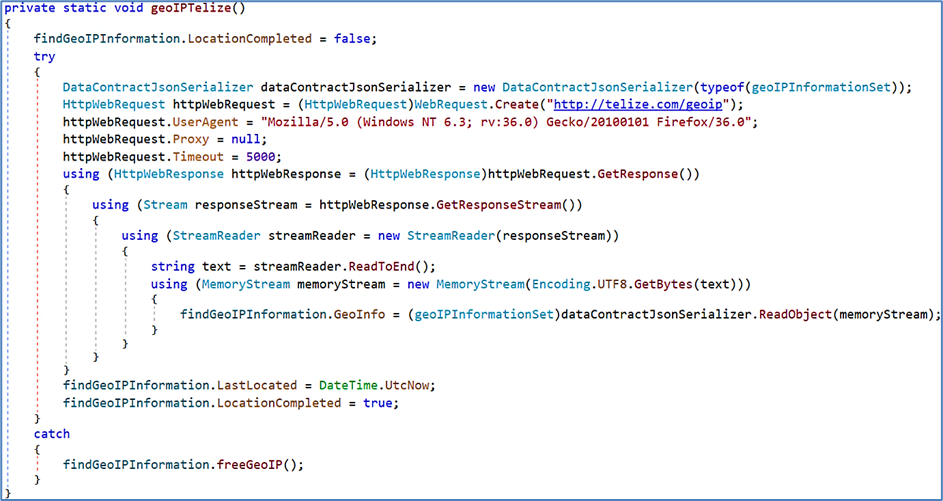}}
    \caption{Interface to query \textit{Telize geoip} function and retrieve victim's location.}
    \label{fig:query}
\end{figure}

At this stage, the function attempts to determine the user's geographical location using the \textit{Telize} service (Fig. \ref{fig:flag} and Fig. \ref{fig:query}). Additionally, it utilizes a list of flag image names to match the identified nationality with the corresponding image string (Fig. \ref{fig:query}).

\begin{figure}[H]
    \centering
    \frame{\includegraphics[width=0.8\linewidth]{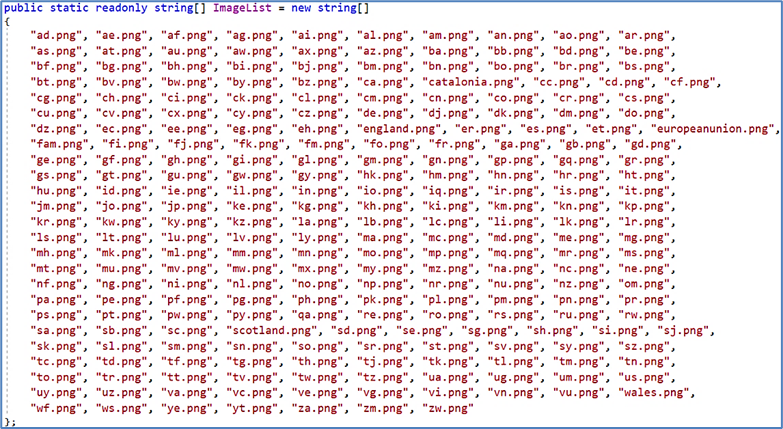}}
    \caption{National flags images list}
    \label{fig:national}
\end{figure}

\begin{figure}[H]
    \centering
    \frame{\includegraphics[width=0.8\linewidth]{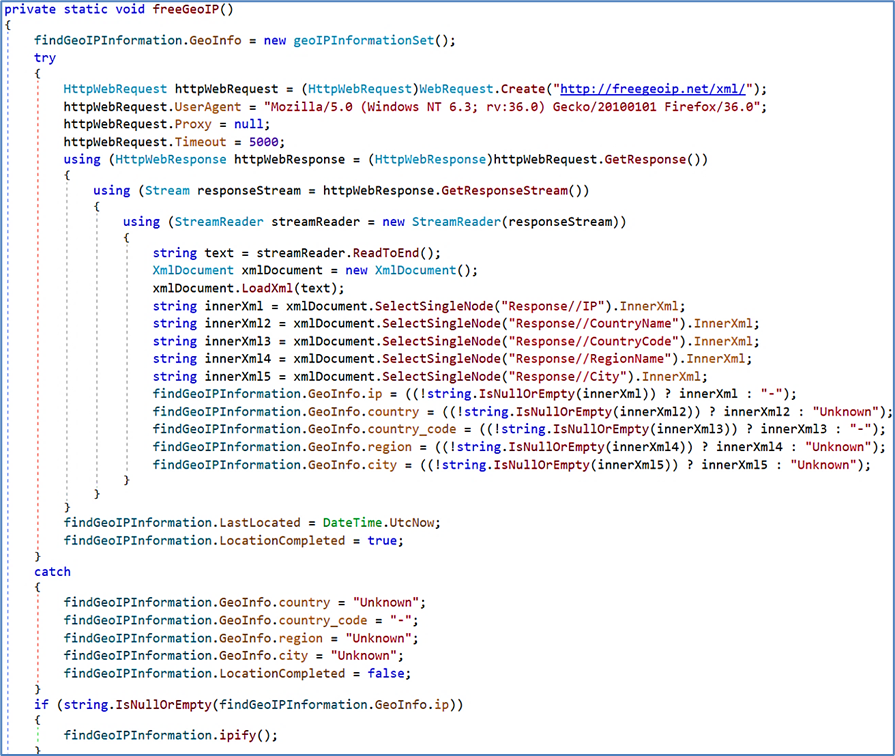}}
    \caption{Function used to track user location with \textit{FreeGeoIP} service.}
    \label{fig:freegeo}
\end{figure}

\begin{figure}[H]
    \centering
    \frame{\includegraphics[width=0.8\linewidth]{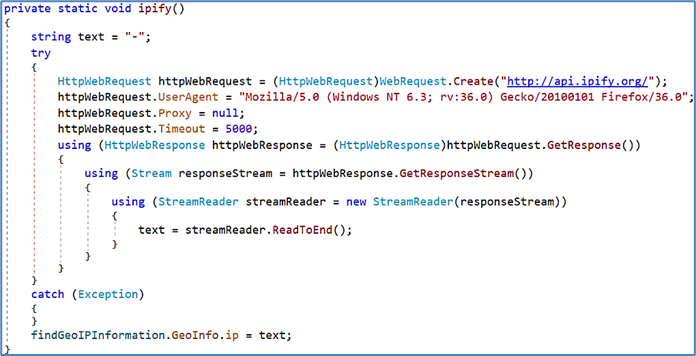}}
    \caption{Function used to track user location with \textit{ipify} service.}
    \label{fig:ipify}
\end{figure}

Additionally, the module implements a double fail-safe mechanism for user geo-localization. If the initial attempt using \textit{Telize} fails, it will revert to the \textit{freeGeoIP} service (Fig. \ref{fig:freegeo}). Should this also fail, a final attempt is made using the \textit{ipify APIs} (Fig. \ref{fig:ipify}).

\begin{figure}[H]
    \centering
    \frame{\includegraphics[width=0.8\linewidth]{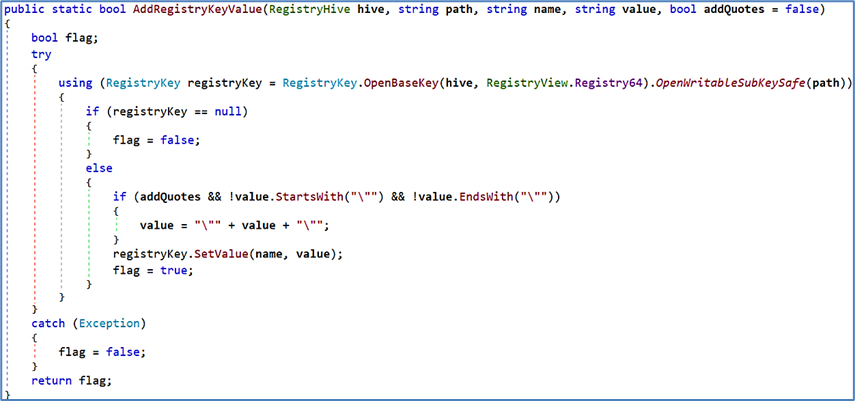}}
    \caption{Primitive used to create a new key inside Windows registry.}
    \label{fig:crkey}
\end{figure}

\begin{figure}[H]
    \centering
    \frame{\includegraphics[width=0.8\linewidth]{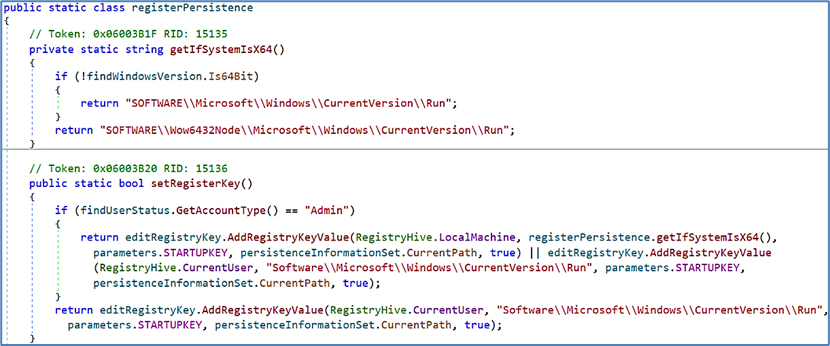}}
    \caption{Function used to set a key inside Windows registry.}
    \label{fig:addkey}
\end{figure}

After determining the victim’s location, the analyzed function attempts to establish persistence by creating a new registry entry under either the \textit{CurrentVersion\textbackslash Run} path or its \textit{Wow6432Node} equivalent (Fig. \ref{fig:crkey} and Fig. \ref{fig:addkey}). The entry’s key name is derived from a value initialized in the parameters class and includes the current path of the \textit{DLL}.

\begin{figure}[H]
    \centering
    \frame{\includegraphics[width=1\linewidth]{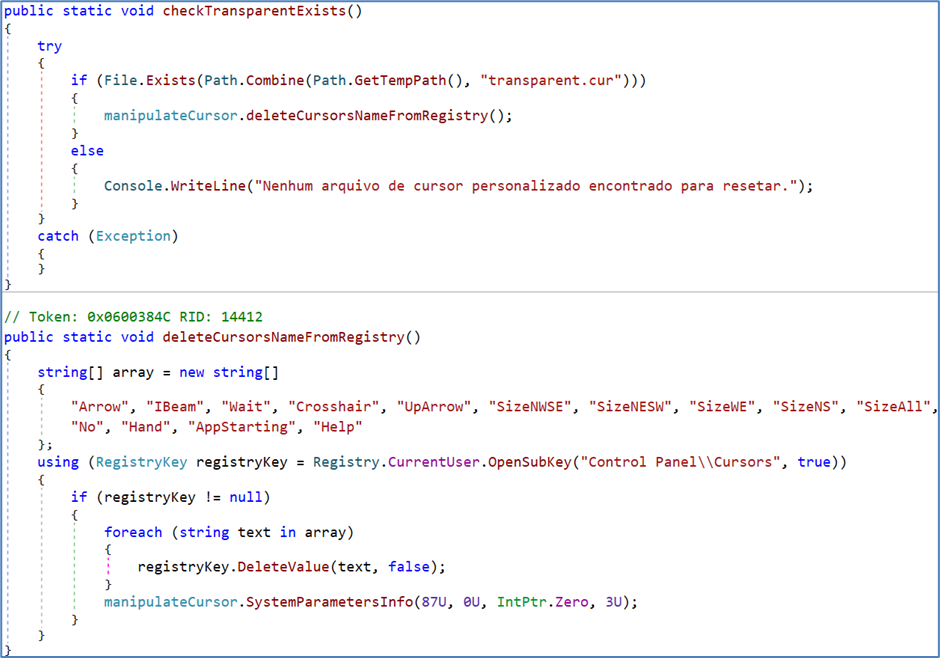}}
    \caption{Function used to check for existence of \textit{transparent.cur} and removal of values from mouse customization reg keys.}
    \label{fig:transp}
\end{figure}

\textit{persistenceAndGeoLoc} function will then first verify the existence of the file \textit{transparent.cur} in the \textit{Temp} directory. If this file is not found, it will proceed to clean up specific hardcoded values associated with mouse customization, particularly within the \textit{Cursor registry subkey}. This cleanup operation will occur only if the \textit{Cursor subkey} contains any entries. Essentially, this function aims to hide the mouse cursor by removing its customization settings, ensuring that the cursor remains invisible to the user (Fig. \ref{fig:transp}).

\begin{figure}[H]
    \centering
    \frame{\includegraphics[width=0.9\linewidth]{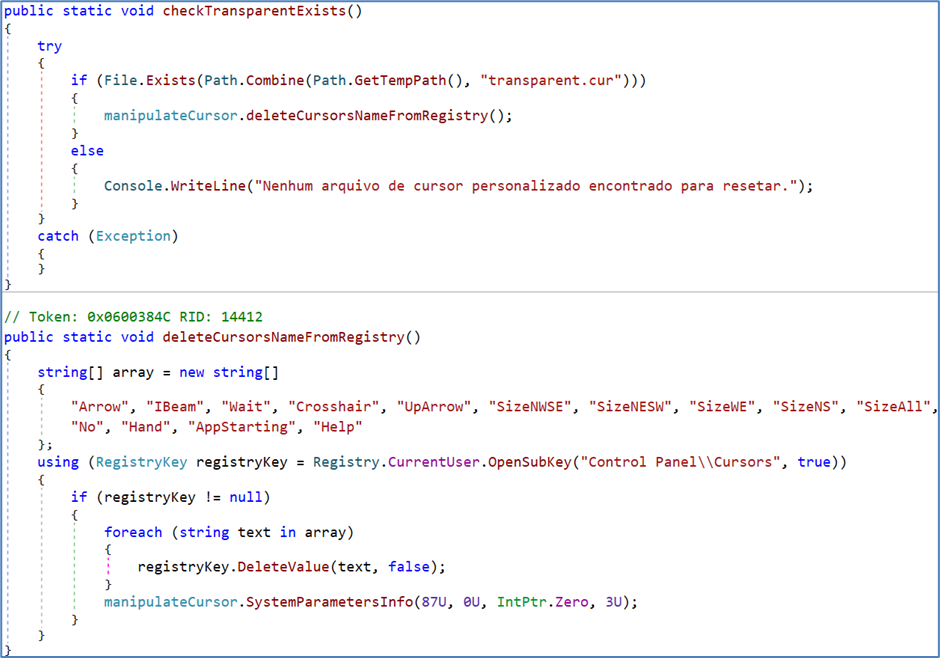}}
    \caption{First half of the function used to install it-self inside victim’s filesystem.}
    \label{fig:fs}
\end{figure}

Finally, the function handles the installation of itself into the filesystem. It begins by checking if the directory \textit{C:\textbackslash Users\textbackslash $<$username$>$\textbackslash AppData\textbackslash Roaming\textbackslash SUB\textbackslash INSTALL} exists. If it does not, the function creates this directory. If files are already present in this location, the function removes them, including terminating any processes that have handles open to these files to facilitate their removal. This process involves an additional delay of \textit{5000 ms} if a file in the \textit{INSTALL} folder is currently in use (Fig. \ref{fig:fs}). After ensuring that the persistence key is properly added to the system registry, the function sets the hidden attribute on the newly copied instance of itself in the \textit{InstallPath}. Finally, it initiates the persistent copy as a new process and disconnects from the \textit{Command \& Control} server.

\begin{figure}[H]
    \centering
    \frame{\includegraphics[width=0.9\linewidth]{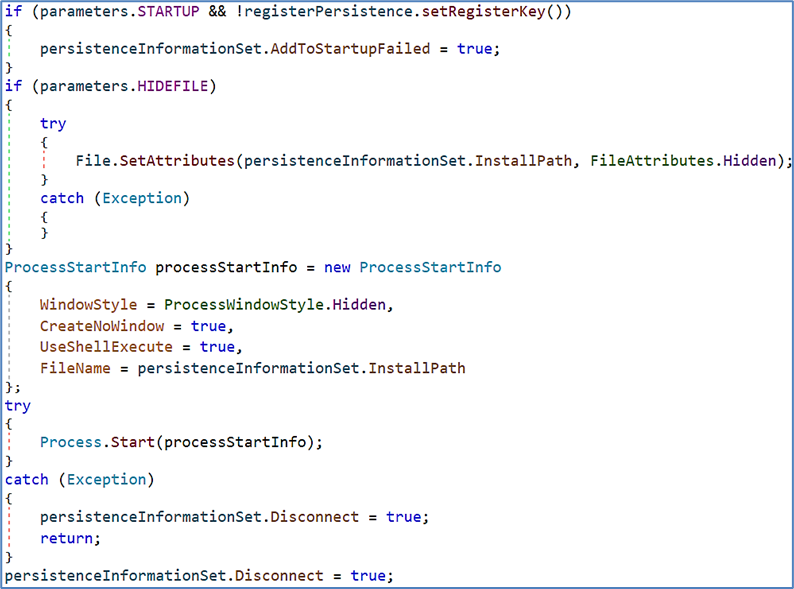}}
    \caption{Second half of the function used to install it-self inside victim’s filesystem.}
    \label{fig:enter-label}
\end{figure}

\begin{figure}[H]
    \centering
    \includegraphics[width=1\linewidth]{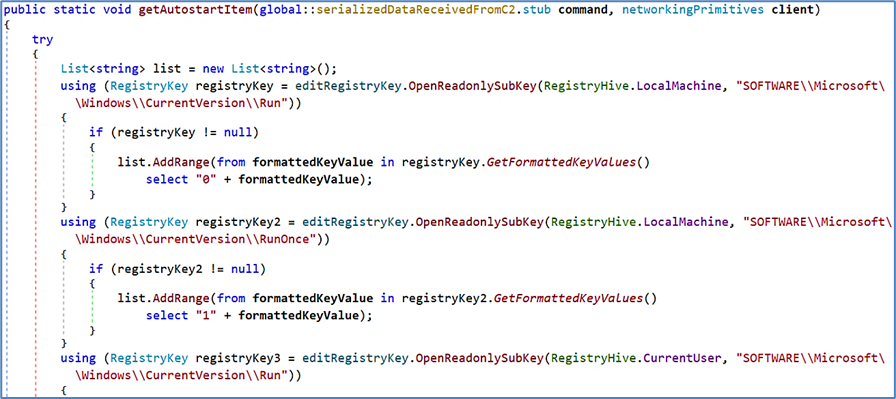}
    \caption{This threat can set, get and delete programs added to the startup registry keys.}
    \label{fig:editreg}
\end{figure}

In addition to the previously mentioned persistence mechanisms, this threat further establishes its foothold by adding new registry keys within the following locations, ensuring that any additional payloads or malicious components are executed each time the user logs on (Fig. \ref{fig:editreg}):

\begin{itemize}
    \item HKEY\_LOCAL\_MACHINE\textbackslash Software\textbackslash Microsoft\textbackslash Windows\textbackslash CurrentVersion\textbackslash Run
    \item HKEY\_CURRENT\_USER\textbackslash Software\textbackslash Microsoft\textbackslash Windows\textbackslash CurrentVersion\textbackslash Run
    \item HKEY\_CURRENT\_USER\textbackslash Software\textbackslash Microsoft\textbackslash Windows\textbackslash CurrentVersion\textbackslash\newline RunOnce
    \item HKEY\_LOCAL\_MACHINE\textbackslash SOFTWARE\textbackslash Wow6432Node\textbackslash Microsoft\textbackslash Windows\textbackslash\newline Run (only for x64 systems)
    \item HKEY\_LOCAL\_MACHINE\textbackslash SOFTWARE\textbackslash Wow6432Node\textbackslash Microsoft\textbackslash Windows\textbackslash\newline RunOnce (only for x64 systems)
\end{itemize}

\begin{figure}[H]
    \centering
    \includegraphics[width=1\linewidth]{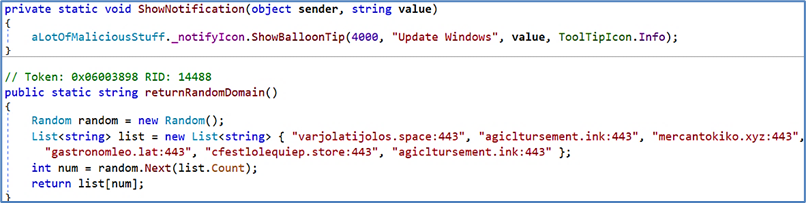}
    \caption{\textit{ShowNotification} and \textit{returnRandomDomain} functions}
    \label{fig:shownot}
\end{figure}

\begin{figure}[H]
    \centering
    \includegraphics[width=0.9\linewidth]{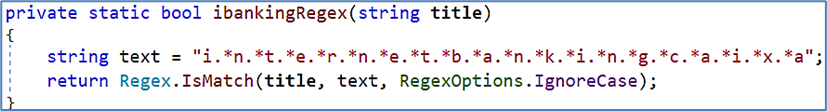}
    \caption{\textit{ibankingregex} function}
    \label{fig:regex}
\end{figure}

Further investigation into the \textit{aLotofMaliciousStuff} class has revealed the existence of three methods that appear to be unused in the current implementation:
\begin{itemize}
    \item \textit{ShowNotification}: This method creates an informational balloon tip with the title \textit{Update Windows} and a custom message body. It is then possible to assume, with good confidence, that this might be used to deceive the user into downloading third-party malicious software. S possible strategy employed by the \textit{Threat Actor} to sell advertising space within their malware, allowing other threat actors to leverage this platform for their own distribution (Fig. \ref{fig:shownot});
    \item \textit{returnRandomDomain}: This function randomly selects and returns one of six domains associated with port 443. The purpose of this method could be to provide backup \textit{Command and Control} servers or to check connectivity to different domains, ensuring resilience and flexibility in communication with the malware's command infrastructure (Fig. \ref{fig:shownot}).
    \item \textit{ibankingRegex}: This method performs a comparison of a given string against a predefined regular expression. A string must contain specific letters, with the possibility of other arbitrary strings between them, to match the regex. This functionality could be used to obfuscate hardcoded strings and evade static detection mechanisms by refining how sensitive information is handled (Fig. \ref{fig:regex}).
\end{itemize}

The presence of these unused functions suggests several possible scenarios. One hypothesis is that the Threat Actor may be developing multiple versions of this threat, each tailored for different operational contexts. By retaining these methods, they can easily swap, add, or remove functionalities as needed. Alternatively, the presence of these extraneous methods, coupled with unhandled exceptions, debugging strings, inefficient code segments, and deprecated functions (such as the use of \textit{RijndaelManaged}, which is deprecated in \textit{.NET 5.0+} in favor of \textit{AesManaged} or \textit{Aes}), may indicate a rushed development process. This could suggest that the adversary faced time constraints and was unable to optimize or refine the code effectively.

\begin{figure}[H]
    \centering
    \frame{\includegraphics[width=0.7\linewidth]{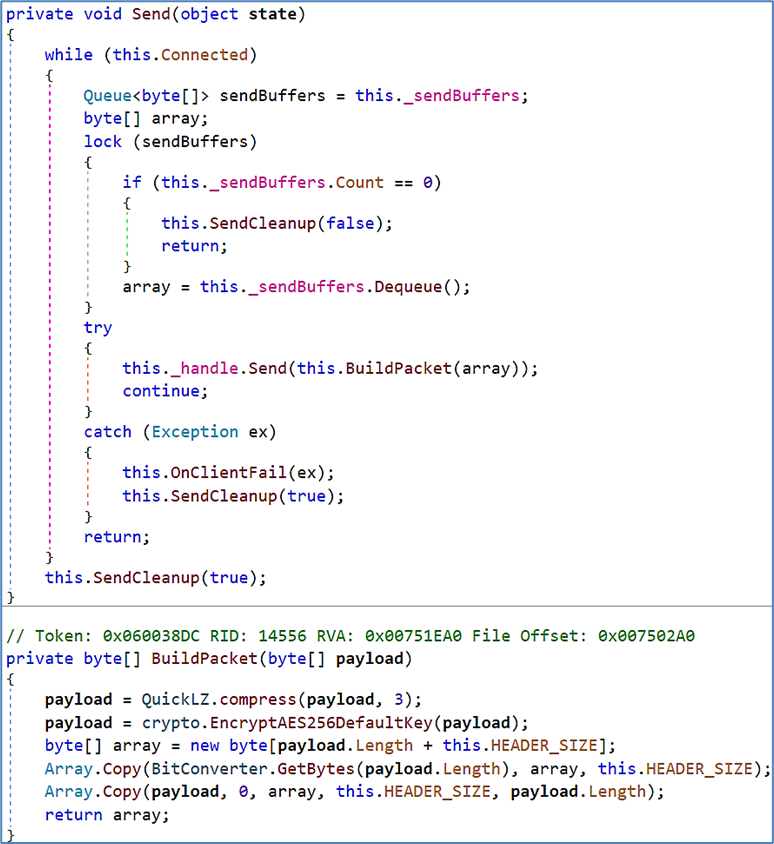}}
    \caption{Function used to prepare and send data to \textit{C\&C}.}
    \label{fig:senddata}
\end{figure}

Examining the threat's packet transmission behavior reveals several key characteristics Fig. \ref{fig:senddata}). Before sending packets over the Internet, it employs a two-stage processing mechanism involving compression and encryption.

\begin{figure}[H]
    \centering
    \frame{\includegraphics[width=0.8\linewidth]{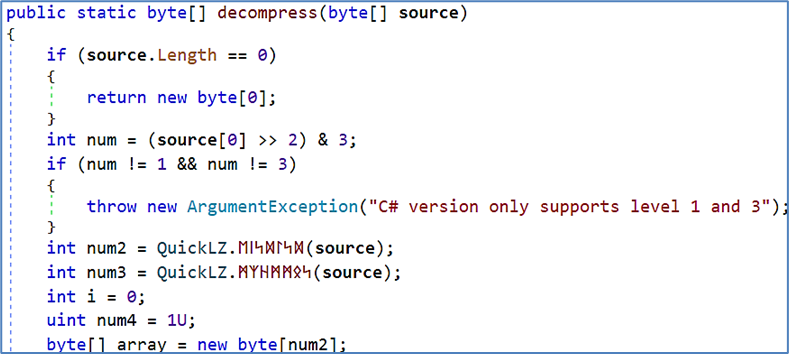}}
    \caption{Function used to decompress \textit{QuickLZ} data shows error handling strings.}
    \label{fig:quicklz}
\end{figure}

\begin{figure}[H]
    \centering
    \frame{\includegraphics[width=0.9\linewidth]{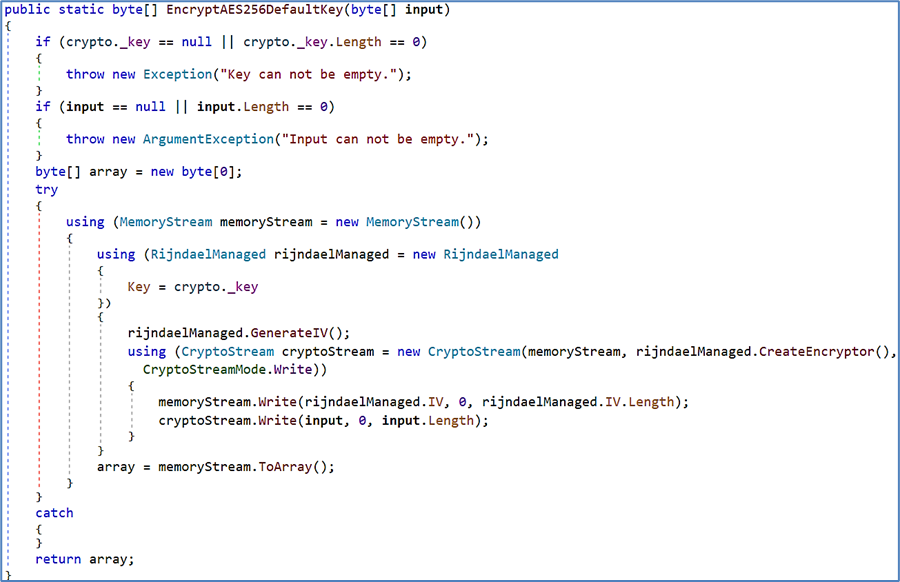}}
    \caption{Function used to encrypt data via \textit{AES256-CBC} with \textit{PKCS7} padding.}
    \label{fig:aes}
\end{figure}

The compression routine, identified through error handling strings such as "\textit{C\#} version only supports level 1 and 3", is associated with the open-source \textit{QuickLZ} compression library (Fig. \ref{fig:quicklz}). This tool is utilized to compress data before encryption. 
The encryption process involves \textit{AES256-CBC} with \textit{PKCS7} padding. In this method, plaintext is encrypted using an \textit{MD5}-hashed version of the \textit{ENCRYPTIONKEY} parameter (Fig. \ref{fig:password}). The resulting encrypted data is a byte array where the first 16 bytes represent a randomly generated \textit{Initialization Vector} (\textit{IV}), and the remaining bytes constitute the ciphertext (Fig. \ref{fig:aes}).

\begin{figure}[H]
    \centering
    \frame{\includegraphics[width=0.9\linewidth]{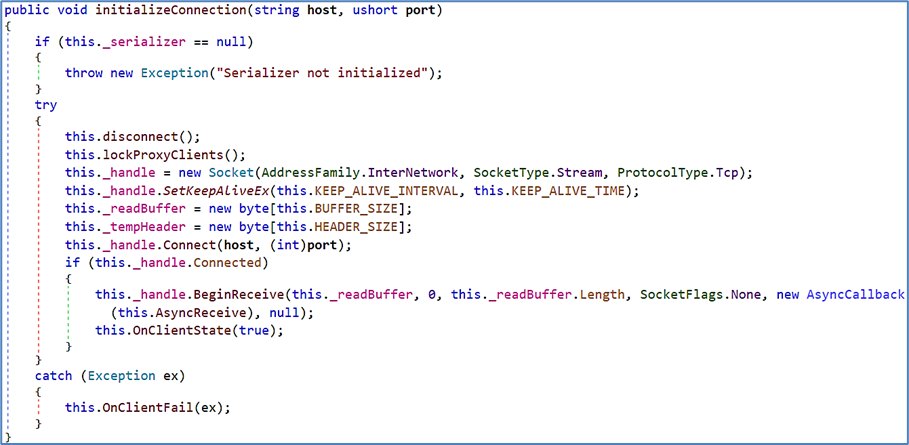}}
    \caption{Setup of the initial connection to the \textit{C\&C} Server.}
    \label{fig:cnc}
\end{figure}

\begin{figure}[H]
    \centering
    \frame{\includegraphics[width=0.65\linewidth]{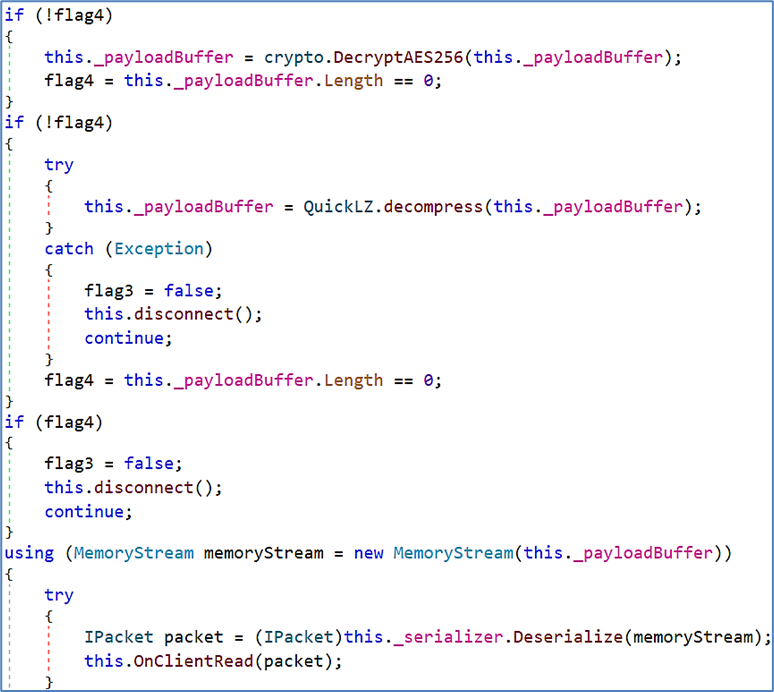}}
    \caption{Main routine of the \textit{AsyncReceive} function which decrypts, decompresses and deserializes \textit{C\&C} received content.}
    \label{fig:async}
\end{figure}

For decryption, the \textit{AsyncReceive} method, which is assigned as the \textit{AsyncCallBack} function in the receiver module of the threat (Fig. \ref{fig:cnc}), performs the inverse operations (Fig. \ref{fig:async}). The process involves decrypting the data, decompressing it, and finally \textit{deserializing} it to retrieve the original content.

These findings underscore the threat's use of robust techniques for data protection and exfiltration, ensuring that transmitted data remains concealed and secure from casual observation.

Within the cryptographic primitives \textit{namespace}, additional unused encryption and decryption functions were identified:
\begin{itemize}
    \item \textit{EncryptAndConvertToB64}: This function is designed to encode data in \textit{Base64} format and then encrypt it using \textit{AES256} with a custom key. Notably, there is no corresponding decryption function for this routine, limiting its utility in the threat’s current implementation. The available decryption method does not support the use of a custom key, relying instead on a standard embedded key;
    \item \textit{EncryptAndConvertToB64DefaultKey}: This method performs \textit{Base64} encoding of data followed by \textit{AES256} encryption using a default embedded key;
    \item \textit{DecryptAES256FromB64DefaultKey}: This function decodes a \textit{Base64}-encoded string and subsequently decrypts it using \textit{AES256} with the default embedded key.
\end{itemize}

\begin{figure}[H]
    \centering
    \frame{\includegraphics[width=1\linewidth]{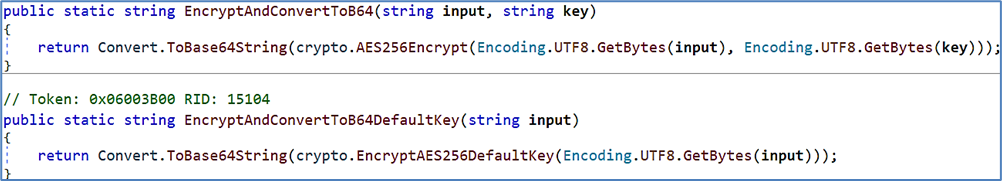}}
    \caption{Unused encoding and encryption functions.}
    \label{fig:encdec}
\end{figure}

\begin{figure}[H]
    \centering
    \frame{\includegraphics[width=1\linewidth]{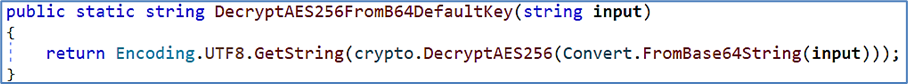}}
    \caption{Unused decoding and decryption function.}
    \label{fig:decdec}
\end{figure}

The presence of these functions, despite their lack of use in the current threat operations, suggests that the threat actor may have intended to implement additional encryption mechanisms or had considered different encryption strategies during development (Fig. \ref{fig:encdec} and Fig. \ref{fig:decdec}).

\newpage
\section{Conclusion}
In conclusion, the malspam campaign deploying \textit{BlotchyQuasar} exemplifies a sophisticated and multifaceted threat operation specifically tailored for \textit{Latin American} targets. The attack's success hinges on its effective use of DLL side-loading, a technique that allows it to execute a wide range of malicious activities while evading traditional security defenses.

This analysis reveals a notable paradox in the threat actor's tradecraft: the campaign pairs a sophisticated and effective attack chain with a codebase that shows clear signs of being rushed, inefficient, and poorly managed. This contrast suggests an adversary who prioritizes rapid, functional deployment over stealth and code optimization. While these flaws may present opportunities for detection and mitigation, they do not diminish the significant harm posed by the malware's potent capabilities for credential theft and system control. Ultimately, this campaign serves as a critical reminder of the evolving threat landscape and reinforces the urgent need for heightened cybersecurity awareness and strengthened defenses, particularly within the financial sector in Latin America where such targeted attacks are on the rise.
\newpage
\appendix
\section{Appendix}
\subsection{IoCs, TTPs \& Yara Rules} \label{App:IoC}
The entire set of \textit{IoCs}, \textit{TTPs} and few \textit{Yara} Rules, gathered through-out this entire analysis, are available inside the following \textit{AlienVault OTX} \href{https://otx.alienvault.com/pulse/685e9d51e6bd9712ae30941b}{pulse}.

\begin{figure}[H]
    \centering
    \includegraphics[width=1\linewidth,frame]{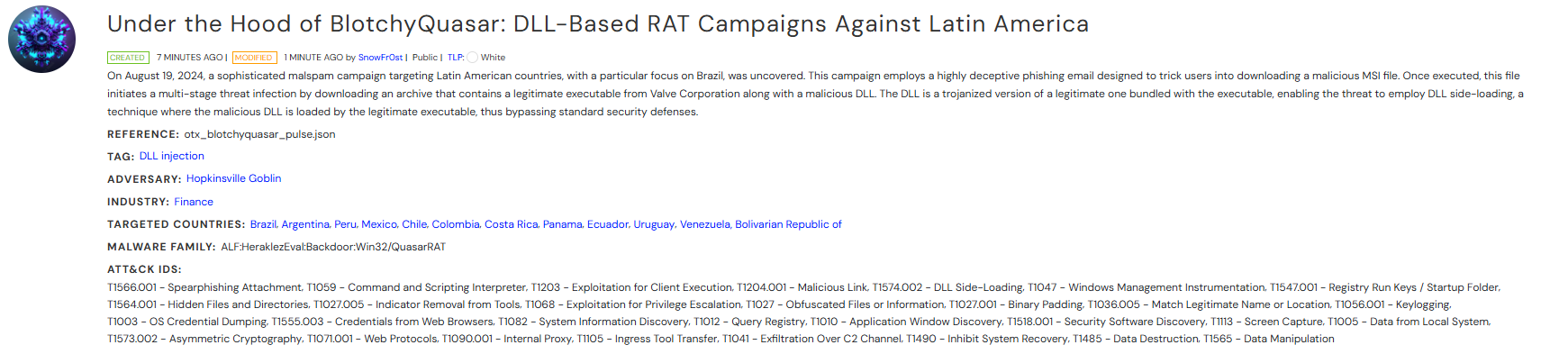}
    \caption{Overview of the \textit{AlienVault OTX pulse}}
    \label{fig:165}
\end{figure}

\newpage

\subsection{Sigma Rules}
\vspace*{\fill} 
\begin{center}
\begin{lstlisting}[language=yaml]
title: Detection of Suspicious AnyDesk File Modification and Termination via PowerShell
id: 1234abcd-5678-efgh-ijkl-9012mnopqrst
description: Detects suspicious PowerShell activity involving AnyDesk file modification and process termination when specific command patterns are observed.
status: experimental
author: Alessio Di Santo
date: 2024-11-26
logsource:
  category: process_creation
  product: windows
detection:
  selection:
    Image: '*\powershell.exe'
    CommandLine|all:
      - 'ad.anynet.pwd_hash='
      - 'ad.anynet.pwd_salt='
      - 'ad.anynet.token_salt='
      - 'taskkill /IM anydesk.exe /F'
  condition: selection
fields:
  - CommandLine
  - ParentCommandLine
  - ParentImage
  - Image
  - User
level: high
tags:
  - attack.persistence
  - attack.t1562.001
  - attack.t1098
falsepositives:
  - Legitimate administrative maintenance involving AnyDesk
mitre:
  - T1562.001
  - T1098
\end{lstlisting}
\end{center}
\vspace*{\fill} 
\newpage
\vspace*{\fill} 
\begin{center}
\begin{lstlisting} [language=yaml]
title: Suspicious MSI Installer Downloaded via Malspam
id: 9fc2e660-1b1d-4b4a-8aaf-6a2b3c71d2f9
description: Detect execution of the malicious MSI installer (67dee1.msi) used in the LATAM BlotchyQuasar campaign.
author: Alessio Di Santo
date: 2025/06/27
tags:
  - attack.initial_access
  - attack.t1566.001
  - malware
logsource:
  product: windows
  service: sysmon
  definition: 'Sysmon Event ID 1: Process Create'
detection:
  selection:
    EventID: 1
    Image|endswith: '\msiexec.exe'
    CommandLine|contains:
      - '67dee1.msi'
  condition: selection
level: high
\end{lstlisting}
\end{center}
\vspace*{\fill} 
\newpage
\vspace*{\fill} 
\begin{center}
\begin{lstlisting}[language=yaml]
title: DLL Side-Loading of vstdlib_s64.dll by SteamErrorReporter.exe
id: 5a7b4282-34f2-4d6b-8c3e-0d6af8f5c6fa
description: Detect when steamerrorreporter.exe loads vstdlib_s64.dll (quasar RAT) via DLL side-loading.
author: Alessio Di Santo
date: 2025/06/27
tags:
  - attack.t1574.002
  - attack.execution
  - malware
logsource:
  product: windows
  service: sysmon
  definition: 'Sysmon Event ID 7: Image loaded'
detection:
  selection:
    EventID: 7
    ImageLoaded|endswith: '\vstdlib_s64.dll'
    ParentImage|endswith: '\EIUWI383IE.exe'
  condition: selection
level: critical
\end{lstlisting}
\end{center}
\vspace*{\fill} 
\newpage
\vspace*{\fill} 
\begin{center}
\begin{lstlisting}[language=yaml]
title: BlotchyQuasar Network Chain: Drop + GeoIP Services
id: 2c8fa042-a80f-4c9a-aba4-3f5d2e08e5d2-mod
description: Detect a connection to at least one drop server domain plus all three GeoIP services used by BlotchyQuasar.
author: Alessio Di Santo
date: 2025/06/27
tags:
  - attack.t1105
  - attack.t1071.001
  - attack.t1573.002
logsource:
  product: firewall
  service: any
detection:
  # At least one drop or staging domain
  selection_drop:
    DestinationHostname|matches:
      - 'notificacao.noticiasnovidads.xyz'
      - 'liga-730ce-default-rtdb.europe-west1.firebasedatabase.app'

  # All three GeoIP lookup services
  selection_geoip_telize:
    DestinationHostname|contains: 'telize.com'
  selection_geoip_freegeoip:
    DestinationHostname|contains: 'freegeoip.net'
  selection_geoip_ipify:
    DestinationHostname|contains: 'api.ipify.org'

  condition: selection_drop and selection_geoip_telize and selection_geoip_freegeoip and selection_geoip_ipify
level: high
\end{lstlisting}
\end{center}
\vspace*{\fill} 

\newpage
\subsection{Infection Chain}
\begin{figure}[H]
    \centering
    \includegraphics[width=0.95\textheight, angle=270,frame]{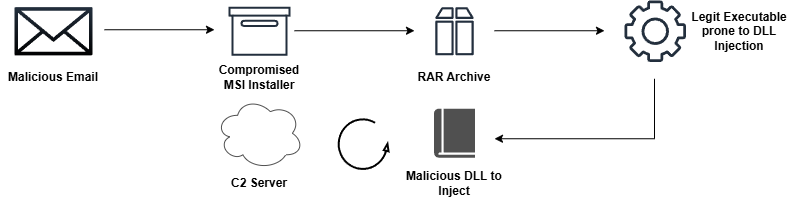} 
\end{figure}

\subsection{Diamond Model}
\begin{figure}[H]
    \centering
    \includegraphics[width=0.95\textheight, angle=270,frame]{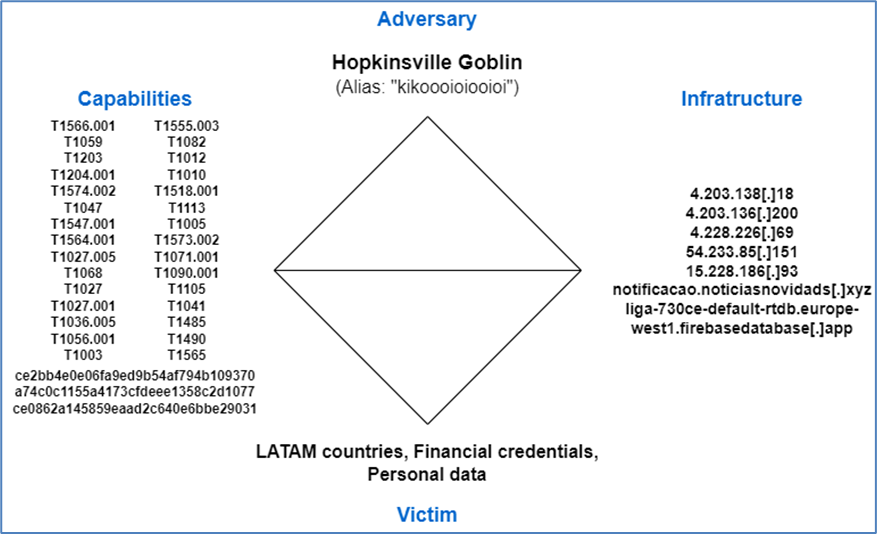}
    \label{fig:DM}
\end{figure}

\end{document}